# Kepler K2 Campaign 9: I. Candidate short-duration events from the first space-based survey for planetary microlensing

I. McDonald,[1,2]* E. Kerins,[1] R. Poleski,[3] M.T. Penny,[4] D. Specht,[1] S. Mao,[5,6] P. Fouqué,[7,8] W. Zhu[5,9] and W. Zang[5]

[1] Jodrell Bank Centre for Astrophysics, Alan Turing Building, University of Manchester, Manchester, M13 9PL, UK
[2] Open University, Walton Hall, Kents Hill, Milton Keynes, MK7 6AA, UK
[3] Astronomical Observatory, University of Warsaw, Al. Ujazdowskie 4, 00-478 Warszawa, Poland
[4] Louisiana State University, 261-B Nicholson Hall, Tower Dr., Baton Rouge, LA 70803-4001, USA
[5] Department of Astronomy, Tsinghua University, Beijing 100084, China
[6] National Astronomical Observatories, Chinese Academy of Sciences, Beijing 100101, China
[7] Université de Toulouse, UPS-OMP, IRAP, Toulouse, France
[8] CFHT Corporation, 65-1238 Mamalahoa Hwy, Kamuela, Hawaii 96743, USA
[9] Canadian Institute for Theoretical Astrophysics, University of Toronto, 60 St George Street, Toronto, ON M5S 3H8, Canada

## ABSTRACT

We present the first short-duration candidate microlensing events from the *Kepler K2* mission. From late April to early July 2016, Campaign 9 of *K2* obtained high temporal cadence observations over a 3.7 deg$^2$ region of the Galactic bulge. Its primary objectives were to look for evidence of a free-floating planet (FFP) population using microlensing, and demonstrate the feasibility of space-based planetary microlensing surveys. Though *Kepler K2* is far from optimal for microlensing, the recently developed MCPM photometric pipeline enables us to identify and model microlensing events. We describe our blind event-selection pipeline in detail and use it to recover 22 short-duration events with effective timescales $t_{eff} < 10$ days previously announced by the OGLE and KMTNet ground-based surveys. We also announce five new candidate events. One of these is a caustic-crossing binary event, consistent with a bound planet and modelled as such in a companion study. The other four have very short durations ($t_{eff} < 0.1$ days) typical of an Earth-mass FFP population. Whilst *Kepler* was not designed for crowded-field photometry, the *K2*C9 dataset clearly demonstrates the feasibility of conducting blind space-based microlensing surveys towards the Galactic bulge.

## 1 INTRODUCTION

Free-floating planetary-mass objects (FFPs) may represent the end-states of disrupted exoplanetary systems (e.g. Rasio & Ford 1996) and "failed" stars (Whitworth & Zinnecker 2004; Gahm et al. 2007). Except for the very young and most massive planets (e.g. Lucas & Roche 2000; Bihain et al. 2009; Quanz et al. 2010; Peña Ramírez et al. 2016; Schneider et al. 2016), such objects are too faint to be detected directly, and their identification relies on gravitational microlensing (e.g. Han et al. 2004; Mróz et al. 2017), whereby the planet traverses the observer's line-of-sight toward a background star. The background star is brightened by a Paczyński (1986) curve, with a characteristic timescale ($t_E = R_E/v_{rel}$), defined by an Einstein (1936) radius ($R_E$) and a relative source–lens tangential velocity ($v_{rel}$). $R_E$ depends on the lens mass, and the relative separations of observer, lens and source (Section

3.2). As $v_{rel}$ and lens distance are generally unknown, FFP candidates are observationally defined by events with short $t_E$, unless second-order effects can be quantified, frequently from simultaneous ground- and space-based measurements (e.g., detection of parallax effects, relative velocities inferred from finite-source effects, etc.; Han 2006; Di Stefano 2012; Udalski et al. 2015b; Henderson & Shvartzvald 2016). Contaminants, including high-velocity stars and planets in wide orbits, can enter the dataset (Di Stefano 2012; Clanton & Gaudi 2017). Thus, FFP abundances are normally calculated on a statistical basis, as an excess number of events with short $t_E$.

Ejected high-mass (gas-giant) planets can only account for a relatively small number of FFPs (Veras & Raymond 2012; Ma et al. 2016), hence an abundance of low-mass lenses would indicate the stellar initial-mass function turns up at masses below the brown dwarf "desert" (e.g. Halbwachs et al. 2000). FFP abundances have been variously estimated as twice (Sumi et al. 2011) or less than a quarter (Mróz et al. 2017) as common as main-sequence stars. One source of tension between the two results arises from differing treatments of faint, highly blended events, where the lensed star contributes less than 10 per cent of the incident flux on that part of the imaging detector: Mróz et al. (2017) claim that the $t_E$ of such events cannot be reproduced satisfactorily, leading to an artificial excess of short-timescale events. Such events may be characterised by the alternative measure $t_{eff} = t_E u_0$ (where $u_0$ is the impact parameter in terms of the angular Einstein radius, $\theta_E$; e.g., Gould 1996), or by the full-width at half-maximum, $t_{FWHM}$. Without knowing the baseline flux coming from the lensed star, confident detection of FFPs is difficult.

* E-mail: iain.mcdonald-2@manchester.ac.uk

*K2*, the "second-light" mission of the *Kepler* satellite (Borucki et al. 2010; Howell et al. 2014), was pointed near Baade's Window in the Galactic Bulge during its Campaign 9 (hereafter *K2*C9; Henderson et al. 2016). The area was simultaneously covered by ground-based surveys (Section 3.1). The large stellar column density predicts many detectable FFPs in this direction (Gould & Horne 2013; Ban et al. 2016). If the FFP population follows a distribution like Sumi et al. (2011), several (1.1–6.3) events due to FFPs were expected to be detectable (Penny et al. 2017), otherwise we expect not to observe any FFPs (<1 events) during the *K2*C9 campaign.

To date, there has been no attempt to blindly search for short-timescale events in the *K2*C9 data. Since, $R_E$ for FFPs may be ≪1 AU (cf. the 0.1–0.8 AU separation between Earth and *Kepler*; Henderson et al. 2016), ground-based observations of *K2*C9 events may obtain very different magnifications to *K2*. Events detectable from one may not be detectable by the other, the microlensing parallax effect can be measured for FFP events (see, e.g., Penny et al. 2017, Figs. 2 & 3), and the degeneracy broken between $v_{rel}$ and $\theta_E$ to better define the mass of the lensing object.

This paper details our blind search. Section 2 describes lightcurve[1] extraction from the *K2*C9 data. In Section 3 we flag known candidate microlensing events, providing a control sample for our selection criteria. Sections 4-8 detail our selection pipeline. In Section 8.4 we present the list of known lensing events that pass through our pipeline and in Section 9 we present our final list of new candidate short-duration ($t_{eff} \ll 2$ days) events. In Section 10 we cross-match our new candidates with photometry from ground-based telescopes. Section 11 summarises the properties of our new events and in Section 12 we present our conclusions.

## 2 LIGHT-CURVE EXTRACTION

### 2.1 Data composition

Calibrated images for the *K2*C9 campaign were downloaded from the Mikulski Archive for Space Telescopes[2]. The observations divide into two sub-campaigns: C9a contains 1290 epochs, with a regular cadence of 30 minutes, between JD 2457501.10 and 2457527.44 (2016 Apr 22 to 2016 May 18),

**Table 1.** Extracted light curve regions. Dates are JD − 2457000.

| Region | Extracted dates | Training epochs | |
|--------|-----------------|-----------------|-----------------|
| C9a1 | 501.10–509.87 | 501.10–502.09 | 518.66–527.44 |
| C9a2 | 509.87–518.66 | 501.10–507.68 | 520.85–527.44 |
| C9a3 | 518.66–527.44 | 501.10–509.87 | 526.44–527.44 |
| C9b1 | 531.32–544.90 | 531.32–532.13 | 558.67–572.44 |
| C9b2 | 544.90–558.67 | 531.32–541.46 | 562.11–572.44 |
| C9b3 | 558.67–572.44 | 531.32–544.90 | 571.44–572.44 |

while C9b contains 2022 epochs between JD 2457531.32 and 2457572.44 (2016 May 22 to 2016 Jul 02).

Data-transfer rates meant it was only possible to retrieve a subset of the observed data. This was split between "postage stamps" on known microlensing events (changed between C9a and C9b), and a much larger "superstamp" which, as an unbiased dataset, we analyse here. Located in the high-stellar-density Baade's Window region of the inner galaxy, this superstamp experiences a high rate of short-timescale microlensing events. It comprises roughly 3.7 deg², centred around $\alpha = 18^h$, $\delta = -28$ deg ($l = 2.0$ deg, $b = -2.3$ deg; Henderson et al. 2016), and includes 3 060 000 pixels. Figure 1 shows the spatial extent of the observed field.

Extracting precise lightcurves from the *K2*C9 data is not trivial. *K2*'s pointing was unstable, having lost two of the its four reaction wheels. *K2* was not designed for precision photometry at such high stellar densities: the large pixel scale (4″) leads to an extremely crowded field. The point-spread function (PSF) is under-sampled, and strong intra-pixel variations exist. Both the PSF and intra-pixel response are wavelength-dependent, giving a very complex pixel-response function (PRF) across the broad wavelength response (∼450–900 nm) of the *Kepler* detectors (Bryson et al. 2010; Vorobiev et al. 2018). Lightcurve extraction therefore depends sensitively on the properties of individual pixels, and on the sub-pixel spatial location of stars and objects blended with them.

### 2.2 Lightcurve extraction

Various techniques have used to overcome these issues (e.g Libralato et al. 2016; Zhu et al. 2017). In less-crowded sky areas, the Causal Pixel Model Wang et al. (2016) can remove instrumental trends and spacecraft motion from a subset of pixels, based on a training pixel set. The latter is assumed to have a constant baseline flux, but in *K2*C9 there are multiple stars in every pixel, each drifting in and out of pixels as the telescope moves. Consequently, we use the modified method of Poleski et al. (2019), MCPM, which was developed to work with *K2*C9 data and mitigates, in part, some of these effects.

Spacecraft motion means there is no one-to-one correspondence between physical detector pixels and the sky coordinates, as tracked by MCPM. We therefore extract a grid of lightcurves in right ascension and declination. Grid points were spaced uniformly at $\Delta\delta = 3.6''$ in declination and $\Delta\alpha = 3.6'' \cos\delta$ (which is 3.14–3.21″) in right ascension, ensuring a grid spacing finer than the 4″ pixel scale.

Through experimentation on known events, variable stars, stars near variable stars, and a sample of field stars, we identified a set of suitable input parameters for MCPM. We extract a lightcurve across six target pixels (n_select = 6):

---

this balances the extraction of most of the PSF, allowing recovery of faint candidates, with the chances of blending by bright stars. As almost every object has a known variable star within its six-pixel PSF, we use a large (400-pixel) predictor matrix, with the ten pixels with the largest residual fluxes removed. This reduces effects of nearby variable sources. The L2-regularisation factor was set to $6000$ pixel$^{-1}$.

A critical issue with this data-reduction process is that the extracted lightcurve depends on a training set given to MCMP. This training set assumes a prior photometric model, which we initially take as a flat lightcurve. At each gridpoint, MCMP extracted the lightcurve in six steps, each extracting one-third of the lightcurve for each sub-campaign. Nominally, the training set for each step bookends the extracted set of epochs, but this is not possible at the start and end of each sub-campaign (C9a and C9b). For these situations, the first or last day of the sub-campaign observations is used as appropriate. Thus, events at the very start or end of each sub-campaign will not be well recovered, however we would not accept partial lightcurves of events as microlensing candidates anyway. Table 1 contains a full list of Julian Dates for the extracted regions and training epochs.

This extraction is computationally expensive ($\sim$45 CPU seconds per lightcurve, $\sim$6 CPU years for the entire field). A total of $3\,752\,653$ lightcurves were extracted across all 3312 epochs, totalling $\sim$12.4 billion individual measurements.

## 2.3 Lightcurve filtering

The six sections of the lightcurve were joined in temporal order, using the average flux for the 24-hour period that overlaps between each section. This may reduce sensitivity to short-timescale microlensing events occurring during these periods, and longer-term microlensing events spanning these periods. However, for this trade-off, we obtain a much-reduced amplitude of low-frequency noise.

The light-curve is then high-pass filtered to further reduce low-frequency noise, using a Gaussian kernel with a 5-day characteristic width. This reduces sensitivity to long-timescale events, but should retain FFPs signals, as they have timescales of <2 days. This will also affect our ability to separate FFPs from bound planets on wide orbits: if the lensing star is at too high a $u_0$, it will produce negligible magnification to the lightcurve, but an orbiting planet may still provide strong magnification, resulting in an observation with two unequal components. This is a common problem in the search for FFPs. However, with our low-frequency noise filtering, the stellar signal will be smoothed out, and a bound planet may be interpreted as free floating.

Filtering also affects measures of $t_E$. Consequently, for the majority of this analysis we use either $t_{\rm eff}$ or $t_{\rm FWHM}$ (the full-width at half-maximum flux of the differential lightcurve), as appropriate.

## 3 INITIAL EVENT IDENTIFICATION

### 3.1 Visual inspection of known events

Reduction of this complex dataset requires an *ad hoc* approach candidate selection and filtering. Our design optimises removal of false positives, while retaining lightcurves of the

Optical Gravitational Lensing Experiment (OGLE-IV; Udalski et al. 2015a) microlensing alerts (Udalski 2003)[3]. We reserve events from the Korea Microlensing Telescope Network (KMTNet; Kim et al. 2016)[4] published by Kim et al. (2018b) to blindly test the model's ability to reproduce new candidates. During the analysis of *K2*C9, additional events were announced on KMTNet website[5] and labelled as "possible" events. Notably for this project, these include KMT-2016-BLG-2554 and -2583. For the purposes of chronological consistency, these are not considered as KMTNet detections until Section 8.4, when we identified them as candidates in the *K2* dataset, and totals up until that point do not discuss them except when mentioned explicitly. Events from the Microlensing Observations in Astrophysics (MOA; Sumi et al. 2013)[6] survey were scrutinised for events within the *K2*C9 superstamp, but there are no short-timescale MOA events during the observations that are not already found by OGLE or KMTNet.

Of the 1927 microlensing alerts during the OGLE-IV 2016 campaign, few are both short and occur during *K2*C9. We test our pipeline on events fulfilling three criteria: (1) they peak during the *K2*C9 observations, (2) they have $t_E$ < 4 days (a highly conservative limit for FFPs, allowing the inclusion of brown dwarfs and FFPs with low $\mu$), and (3) they occur within the *K2*C9 superstamp. Visual inspection of the *K2*C9 lightcurves shows that OGLE-2016-BLG-0814 is too complex to expect automatic interpretation as a single-lens, single-source event. A further three (-1041, -1058, and -1110) could not be expected to be recognised by eye without *a priori* knowledge, and two (-1097 and -1162) are only faintly visible and consistent with sporadic noise. Hence, only five lightcurves are expected to be detected (OGLE-2016-BLG-0813, -0878, -1043, -1231, and -1245). We track these, plus events with $t_{\rm eff}$ up to 10 days, through the false-positive rejection process, to minimise deselection of real candidates[7]ote that this set includes OGLE-2016-BLG-0559. This shows two components: one long, faint event, and one short, bright event (cf., Section 2.3). The bright event is modelled by OGLE's Early Warning System as having $t_{\rm eff}$ = 8.2 days, thus meets our criteria for inclusion..

The set of 265 KMTNet events partially overlap with OGLE-IV. Of these, 40 have modelled $t_{\rm eff}$ < 4 days, with many $t_{\rm eff}$ substantially differing from the $t_E$ of their OGLE counterparts. Of these 40, only 30 were visible in the lightcurves, and only 22 have sufficient noise contrast that a by-eye search would have identified them as microlensing events. One (KMT-2016-BLG-0240) peaks outside the *K2*C9 observation window, reducing our expected set of detected candidates numbers only 21.

Note that the $t_E$ and $t_{\rm eff}$ of both OGLE and KMTNet alert systems should be used with caution: they often cannot measured accurately because of degeneracy with $u_0$ and/or blended flux. To circumvent this, the results by OGLE sometimes assume no blending flux, and KMTNet assumes that $u_0$ is either zero or unity (Udalski et al. 2003; Kim et al. 2018a).

---

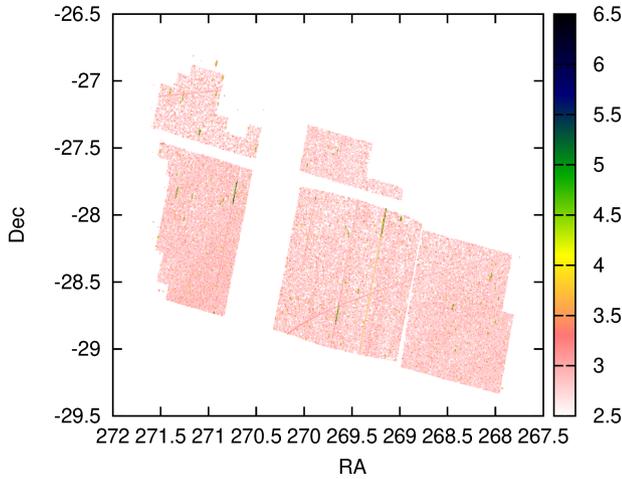

**Figure 1.** Results of initial matched filtering, showing filtered power compared to a Paczynski function with $t_E = 7.6$ hours and $u_0 = 0$. Red colours show events with weak power; blue colours show stronger power. Diagonal stripes from saturated stars and curved trails from asteroid tracks are visible.

## 3.2 Basic matched filtering

Matched filtering (e.g. Helstrom & Wilson 1970) is commonly used to detect template functions in noisy datasets. A simple matched-filter analysis was performed on every lightcurve to obtain a first glance at the false-positive characteristics of the dataset. Our chosen filter is a Paczyński (1986) function, defined as the magnification factor ($A$) expected from a point source undergoing gravitational lensing, as a function of time ($t$), namely

$$A(t) = \frac{u^2 + 2}{u\sqrt{u^2 + 4}},\tag{1}$$

where $u$ is the angular displacement in angular Einstein radii ($\theta_E$) of the lens from the line-of-sight to the source with minimum $u_0$ occurring at $t_0$:

$$u = \sqrt{\left(\frac{t - t_0}{t_E}\right)^2 + u_0^2}.\tag{2}$$

For this initial analysis, we ignore stellar blending.

Each template Paczynski function was first smoothed by a 5-day Gaussian kernel and normalised to an integral of unity, to ensure consistency with the pre-processing of the observations. A variety of $t_E$ and $u_0$ were tested.

Figure 1 maps the results of the $t_E = 10^{-0.5}$ days, $u_0 = 0$ case[8], which typifies a close approach to a source star by an FFP of moderate mass. Time-integrated power in each lightcurve is shown in units of flux (electrons per second, e⁻ s⁻¹) times time (exposures). Figure 1 shows only events with $> 500$ e⁻ s⁻¹ exposures (i.e., $\gtrsim 900\,000$ e⁻) in total differential flux. Most fall into three categories: (1) diagonal, straight lines corresponding to charge-transfer 'bleeding' along detector columns from saturated stars; (2) curved tracks, from passing asteroids; (3) randomly distributed point sources, from a combination of transient events and variable stars. While instructive, these complexities mean we cannot take this matched-filtering approach further at this point, so we return to a more fundamental approach to the data analysis.

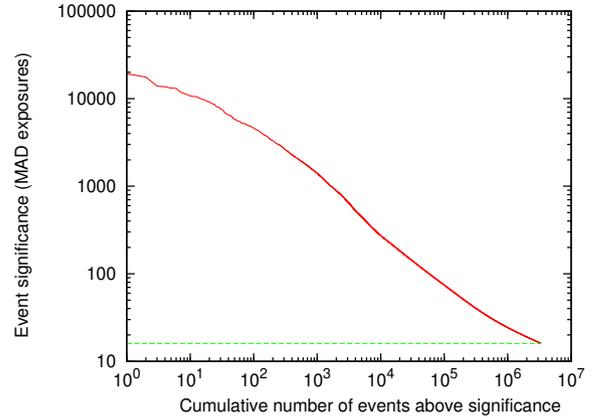

**Figure 2.** Distribution of significance ($s$) among the lightcurves from our initial selection. The horizontal line marks our search limit.

## 3.3 Automated event identification

The complexities of lightcurve extraction, close mimicry of false positives, and lack of a robust training set, mean that typical 'big data' tools (e.g., machine learning) are not suitable for detecting microlensing events in this data. However, the dataset's size requires a comparatively simple solution. Our adopted solution combines combines pre-filtering and pre-whitening of the data with a generalised analysis to detect peaks in the lightcurves.

*Cosmic-ray pre-filtering:* Early removal of photometric spikes caused by cosmic rays led to better recovery of the OGLE test cases. To detect these, the original lightcurve ($F_0$) was convolved with a five-point Hanning function ($H_5$) and differenced from the original lightcurve. The median absolute deviation (MAD) of this differenced lightcurve was calculated, and any noise spikes $>14.826$ MADs were removed[9]. Mathematically expressed, this removes any points where:

$$F_0 > F_0 * H_5 + 14.286\,\text{MAD}\,(F_0 * H_5).\tag{3}$$

Numerically, this is similar to a very mild low-pass filter, but offers a hard cut that prevents downweighting of all but the shortest-timescale microlensing peaks. These ultra-short-timescale ($t_{FWHM} \lesssim 1$ hour) events should be rare, due to finite-source effects.

*Pre-whitening:* Long-timescale variations ("red noise") have various astrophysical and artificial causes. Here, they appear dominated by small-amplitude rotational and pulsational variable stars. To remove these, we now perform a high-pass filter, subtracting a Gaussian-smoothed lightcurve of characteristic width 2 days. This reduces our sensitivity to events with timescales $>2$ days, but timescales for FFPs are normally much shorter (e.g. Mróz et al. 2017).

*Peak identification:* The lightcurve was then scanned for zero-flux-crossing events, between which occur either relative brightening (if positive) or dimming (if negative) compared to the baseline flux. Time-integrated flux in these regions was summed, and significance ($s$, in multiples of MAD over all exposures) calculated. Peaks were accepted if: (1) $s > 16$ MAD exposures, and (2) the peak flux is ranked among the highest 5 per cent of measurements in that lightcurve.

---

[8] The model is quantised, so the singularity at $u = 0$ is not encountered.

[9] The PYTHON `scipy.stats.mad` function was used for this calculation, which employs an automatic scaling factor of 1.4826. Hence, this corresponds to a function result $>10$.

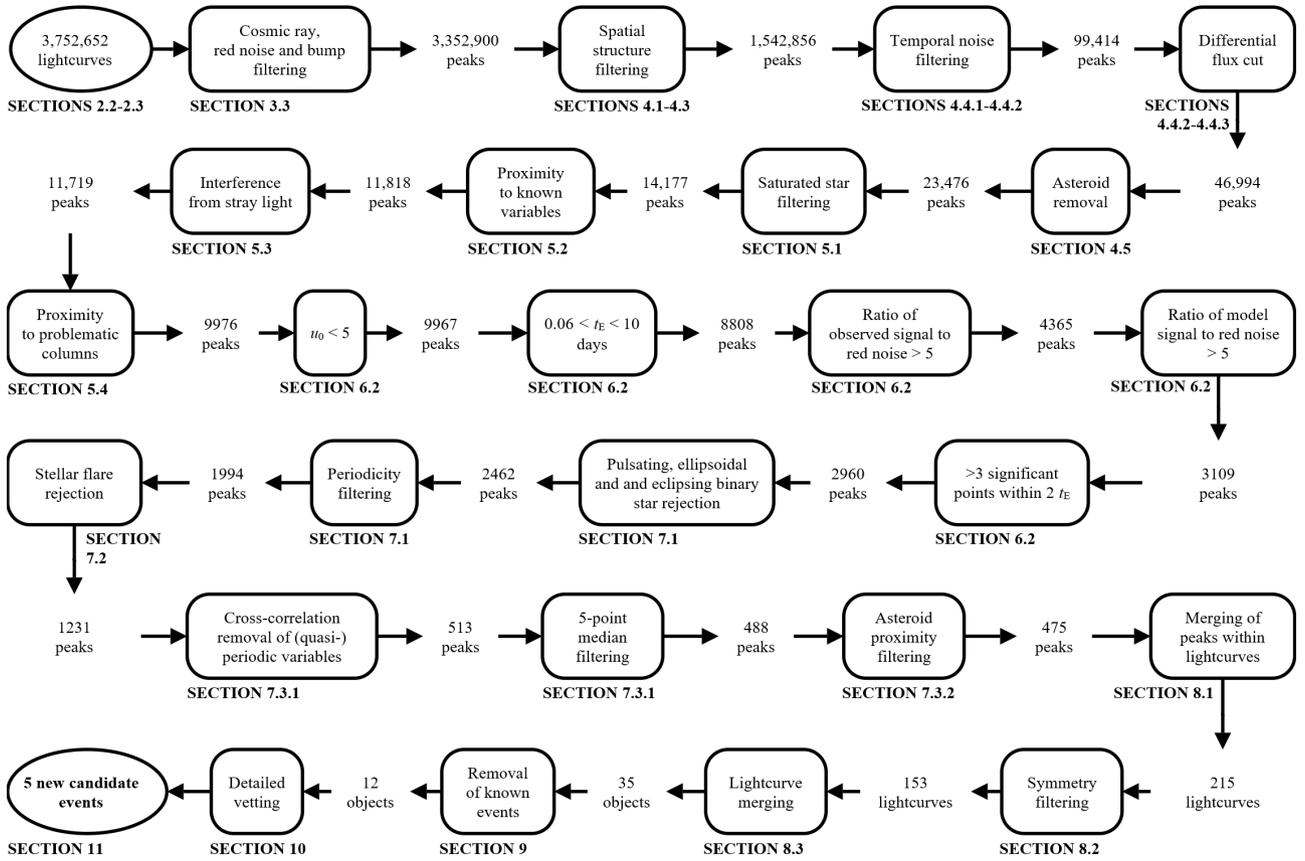

**Figure 3.** Flow diagram, showing the progressive rejection of individual peaks, lightcurves and events, with corresponding section numbers.

This process should robustly detect any abnormal brightening of a star, identifying a large number of peaks that can be filtered to recover viable events. Within $3\,752\,653$ initial lightcurves, $3\,352\,900$ peaks pass this filter. The distribution of $s$ over these peaks is shown in Figure 2.

We compared the extracted events to the known lensing candidates, accepting a match at the position and within $\pm 5$ days as a correctly recovered candidate. This initial stage of filtering matches all five expected OGLE events, plus the faint events OGLE-2016-BLG-1097 and -1058, and 13 of the 19 KMTNet events. The six missed events include KMT-2016-BLG-0180, the complex binary lightcurve; -0231, which occurs at the start of the observations; -0019 and -0024, which suffer contamination by nearby variables; and -0005 and -0068, which suffer from time-dependent noise. Thus, many known events are already missing due to blending.

## 4 RE-WEIGHTING FOR RECURRENT FALSE POSITIVES

### 4.1 Overview

The vast majority of remaining peaks are clearly not microlensing, and must be rejected by a series of cuts, outlined in Figure 3 and subsequent sections. To deal with false positives, we first adopted a series of weights, based on the likelihood that they belonged to each of these criteria. This involved identifying large-scale structures, counting the number of events per spatial co-ordinate ($n_{\alpha\delta}$, flagging persistent effects from saturated or variable stars), and per exposure ($n_t$, flagging "glitches", including spacecraft repointing).

### 4.2 Correlated structure identification

The Einstein radii of microlensing events are orders of magnitude smaller than the telescope PSF, therefore events larger than the PSF will be false positives. To detect coherent spatial structure among surviving peaks, single-linkage clustering in right ascension, declination and time was applied to the data. Single-linkage clustering is optimally successful at identifying long clusters of objects, e.g., both long-term variability and asteroids. Variable stars with periods of $P \lesssim 2$ days can also be isolated using this method, showing up spatially correlated events.

The SCIPY package FCLUSTER was used to perform the clustering, with a characteristic distance of $4''$. A factor of $1''.8$

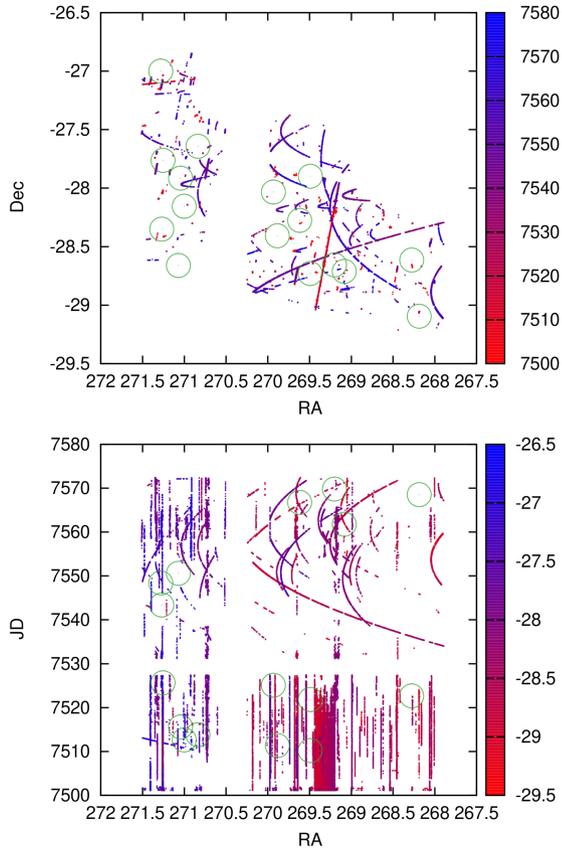

**Figure 4.** Large-scale or temporally repeating structures filtered by single-linkage clustering. Circles indicate OGLE events. Colours represent Julian Date and declination for top and bottom panels, respectively.

per hour ($\sim$1″ per exposure) was used to normalise the time axis, the speed of a slow-moving asteroid. Lightcurves were processed in spatial chunks of $15' \times 15'$ to make the process less memory intensive. The MAD of each cluster's points was calculated in each axis ($r_\alpha$, $r_\delta$ and $r_t$ for RA, declination and time, respectively), and combined to give a characteristic size of the cluster:

$$r_{\rm cl} = \sqrt{\left(\frac{r_\alpha}{\cos\delta}\right)^2 + r_\delta^2 + \left(\frac{r_t}{1.''8/\text{hour}}\right)^2}. \qquad (4)$$

Figure 4 shows those structures with sizes (or size-normalised times) greater than 0.°01, clearly showing many asteroid trails (curved lines in both panels) and variable stars (vertical lines in bottom panel).

### 4.3 Lightcurve weighting

Next, we generated weights for each peak based on their cluster extent ($w_{\rm cl}$), and frequency in space ($w_{\alpha\delta}$) and time ($w_t$). The weighted significance for each lightcurve ($s_{\rm w}$) is given by

$$s_{\rm w} = s\, w_{\rm cl} w_{\alpha\delta} w_t. \qquad (5)$$

Each weight is based on a Gaussian cumulative distribution functions CDF($d$; $c$, $\sigma$), for observed value $d$, mean $c$ and scale

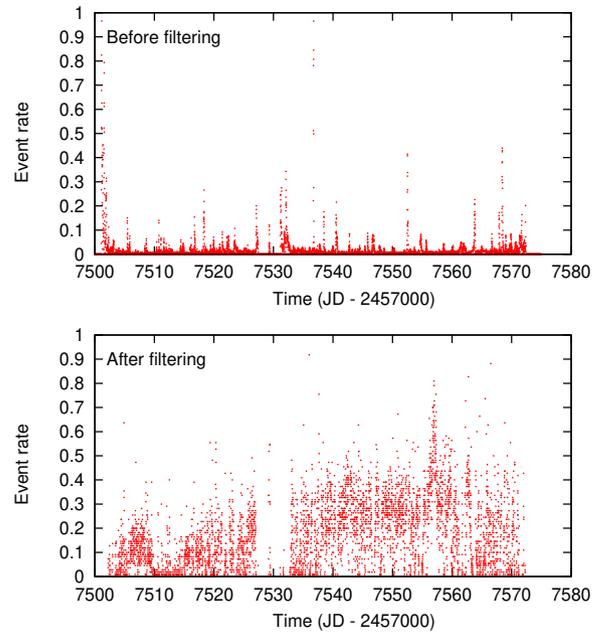

**Figure 5.** Relative rate of *K2C9* peaks before and after filtering. The top panel is used as a noise template ($f_{\rm noise}$) to downweight and remove peaks, resulting in the bottom panel. The spike near JD 2457557.6 is from an unplanned 20″ roll offset, and reversal of the telescope drift rate due to solar radiation pressure.

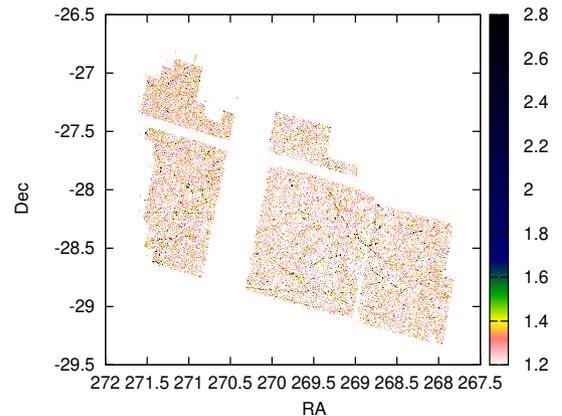

**Figure 6.** Distribution of peaks after weighting. Colours are as in Figure 1.

$\sigma$, softly cutting off large clusters, positions or times:

$$
\begin{aligned}
w_{\rm cl} &= 1 - \text{CDF}(r_{\rm cl}; 0.°01, 0.°005), \qquad (6)\\
w_{\alpha\delta} &= 1 - \text{CDF}(\log_{10} n_{\alpha\delta}; 1.3, 0.3), \text{and}\\
w_t &= 1 - \text{CDF}(\log_{10} n_t; 2.8, 0.2).
\end{aligned}
$$

The mean and scale of each distribution was chosen *a posteriori* based on visual inspection of the retained/rejected peaks and their distribution. These cuts are conservative, minimising false negatives. Only peaks with $s_{\rm w} > 16$ were retained, reducing the 3 352 900 peaks to 1 542 856. All detected OGLE and KMTNet candidates passed this cut.

### 4.4 Saturated and blended stars

#### 4.4.1 Column bleeding

Despite this weighting, visual inspection shows a significant fraction of bright peaks arise from trails from bright, saturated stars that leave sporadic brightening patterns as charge bleeds down the CCD columns. Affected positions change as stars move between and within pixels on the detector, and as the spacecraft rotates. These effects are coherent in time. To address this, we computed two additional weights.

The first weight targets structures aligned with detector columns. Peaks were binned along the detector axis ($\sim 12.2°$ rotation west of north) in $3''6 \times 2$ hour bins, and a weight, $w_{col}$, was assigned based on the number of peaks in that bin ($n_{col}$) following

$$w_{col}(t) = 1 - \text{CDF}(n_{col}; 30, 10). \tag{7}$$

#### 4.4.2 Temporally correlated noise

In addition, many physically isolated but temporally coincident peaks remain. *Kepler*'s under-sampled PSF means that flux from individual stars is distributed over a small number of pixels. When combined with the under-sampled PSF, the high stellar density, and significant spatial differences in sensitivity within each pixel, the blended flux from nearby bright stars can vary significantly as the spacecraft pointing drifts. Such cases appears to represent the vast majority of false positives remaining in the dataset.

Therefore, we used the normalised number of peaks per exposure ($n_t$) to define a pure noise "lightcurve":

$$f_{noise}(t) = \frac{1 + n_t(t)}{1 + n_t^{max}}, \tag{8}$$

where $n_t^{max}$ is the exposure with the most peaks. The top panel of Figure 5 shows the resulting pure-noise lightcurve, indicating most peaks fall within a few specific time bins.

Peaks were further smoothed by an aggressive Gaussian high-pass filter ($\sigma = 0.4$ days), removing structures not associated with noise peaks. The heavily smoothed lightcurves ($f_{sm}$) were compared to the pure-noise lightcurve, and the ratio of "noise-" to "signal-like" flux calculated over all exposures ($t$) using

$$R_{noise} = \sum_t \frac{f_{noise}(t) f_{sm}(t)}{f_{noise}(t)(1 - f_{sm}(t))}. \tag{9}$$

Hence, large $R_{noise}$ indicates events associated with blended stars and spacecraft motion, whereas small $R_{noise}$ indicates a potential event. The mean $R_{noise}$ is 0.0321 and standard deviation is 0.0096. To retain real events (including those scattered by noise up to $\sim$one standard deviation), we create a weight, $w_{noise}$ that downweights noise-like signals:

$$w_{noise} = 1 - \text{CDF}(R_{noise}, 0.0321 + 0.0096, 0.0096), \tag{10}$$

and create a new significance, $s'_w$, such that

$$s'_w = s_w w_{col} w_{noise}. \tag{11}$$

Again, peaks with $s'_w > 16$ were kept, reducing 1 542 856 peaks to 99 414. Their temporal and spatial distribution after filtering are shown in Figure 5 (bottom panel) and Figure 6.

One OGLE candidate (-0878) was partially lost during this cut: its grid point was flagged as correlated noise. However, a neighbouring grid point remained. Five KMTNet candidates were lost (-0051, -0134, -0155, -0206, and -0209) and KMT-2016-BLG-0151 was partially lost. Of these, -0151, -0134, -0151, and -0155 have longer timescales ($\sim$4–6-day) in OGLE and, if such timescale are accurate, would not be expected to survive the vetting process. The two remaining KMTNet candidates (-0206 and -0209) were not detected by OGLE.

**Table 2.** Regions searched for asteroids.

| RA | Dec | Radius |
|---|---|---|
| 17 53 20.00 | −28 45 00 | 40' |
| 17 58 00.00 | −28 00 00 | 1° |
| 18 04 00.00 | −27 48 00 | 1° |

#### 4.4.3 Differential flux weighting

Many remaining peaks have substantial jitter in their wider lightcurves, as the photocentres of bright stars move in and out of the targeted pixels when the spacecraft moves. While many were caught by the $s'_w > 16$ cut, some remain. They are characterised by a sawtooth-shaped lightcurve with a time-variable amplitude, while real events should have a much smoother lightcurve away from the main peak.

The raw lightcurves were again subject to a Gaussian high-pass filter ($\sigma = 2$ days). The standard deviation of this filtered lightcurve is taken, masking out event midpoint and the two days on either side. The masked epochs, representing the event, are then sorted in brightness and the top four epochs taken. Peaks were passed provided the flux in those four events exceeded 5, $^5/_2$, $^5/_3$ and $^5/_4$ times the standard deviation outside the event, respectively. This cut also removed spurious events where only one epoch exceeded the noise threshold, which could be due to instrumental issues, e.g., remnant cosmic rays. The total number of peaks was thus reduced from 99 414 to 46 994. All OGLE and KMTNet candidates passed this cut.

### 4.5 Asteroid removal

Most visually obvious false positives in Figure 6 are from asteroid trails that were not identified as long arcs.

To identify the observed asteroids, three circular apertures were set up to cover the entire *K2C9* field, detailed in Table 2. The JPL/NASA ISPY service[10] was queried in each region between JD 2 457 500 to 2 457 575 in steps of 2.5 days: 1149 asteroids exceeded magnitude 21.0 during any one of these epochs. Ephemerides for each were generated in 30-minute intervals using the HORIZONS service[11].

The Euclidean distance of each event to each ephemeris was computed, and the closest ephemeris chosen. Events were rejected if that distance was <0.07° and the time difference was <0.2 days. *Kepler* detected 477 individual asteroids, and their removal halved the number of peaks from 46 994 to 23 476. All OGLE and KMTNet candidates passed this cut.

---

[10] https://ssd.jpl.nasa.gov/x/ispy.html.
[11] https://ssd.jpl.nasa.gov/horizons.cgi.

## 5 REGIONAL MASKING

Many remaining peaks are heavily down-weighted but exceptionally strong false positives. This section describes the techniques used to entirely mask the worst cases.

### 5.1 Bright stars

A stellar catalogue, covering the regions in Table 2, was extracted from *Gaia* Data Release 2 (*Gaia* DR2; (Gaia Collaboration et al. 2018)), and rotated into detector co-ordinates (see Section 4.4.1).

Masks were generated around *Gaia* objects depending on their $R_p$-band magnitude. Three cuts were made:

- $G_{Rp} < 9$ mag. Around each star, three rectangular regions were masked. The first ($0.0025°$ either side of the star, over the entire detector length) masks complete column bleeds. The second ($\pm 0.0035° \times 0.25°$) masks shorter bleeds from the star's extended PSF. The third ($\pm 0.005° \times 0.05°$) masks effects from the PSF halo.
- $9 \leq G_{Rp} < 10$ mag. A single window of $\pm 0.0035° \times 0.01°$ was masked around stars of to remove shorter column bleeds.
- $10 \leq G_{Rp} < 11$ mag. A single window of $\pm 0.0025° \times 0.005°$ was masked around marginally saturated stars.

While some problematic sources remain, the list of peaks was reduced from 23 476 to 14 177 events. KMT-2016-BLG-0133 and the remaining lightcurve associated with -0151 (OGLE-2016-BLG-0813) were removed by this mask.

### 5.2 Variable stars

A list of 61 941 variable stars was compiled, the vast majority being from the OGLE catalogue of variable stars[12] (Soszyński et al. 2013, 2016), with some additions from other surveys. The sensitivity of OGLE to variable stars varies over the superstamp, as the OGLE-III catalogue only covers the eastern part of the *K2C9* superstamp. Overall, however, most positions in *K2C9* are $\lesssim 18''$ from a known variable star: fully masking every variable star would also mask many chance alignments. We chose instead to mask a smaller aperture, according to three magnitude restrictions: (1) a circle of $3''.6$ ($\sim 1$ *Kepler* pixel) around every variable star; (2) $7''.2$ around stars with $V < 15$ or $I < 15$ mag, or with amplitudes $\Delta I > 1$ mag; (3) $10''.8$ around stars with $V < 12$, $I < 12$, or $\Delta I > 2$ mag. This reduced 14 177 peaks to 11 818. KMT-2016-BLG-0138 and -0149 were masked by this cut. The latter is matched to OGLE event OGLE-2016-BLG-0914, with $t_E \approx 8$ days.

### 5.3 Stray light

A stray light event occurred towards the end of the campaign. The artefacts it created are not visible on the raw images, but show up in the differenced images as straight lines. The underlying cause was not identified, but could be related to the 'Argabrightening' events, thought to be caused by impact debris from the telescope structure (Witteborn et al. 2011). The 99 associated peaks were removed by creating a diagonal mask in the detector plane, removing peaks where $-28.2 <$

$\delta + (\alpha - 271.1)/1.5 < -27.7$ (where $\alpha$ and $\delta$ are the RA and Dec in decimal degrees) and $2457565.7 < \text{JD} < 2457565.8$.

### 5.4 Problematic columns

Strips were created across the detector column axis (cf., Section 4.4.1) of width $0.001°$ by 0.1 days (approximating the PSF radius and shortest recoverable event). Strips containing ten or more events were masked. This primarily affected sections of module 15, where bright the high stellar density led to additional saturation for overlapping bright stars.

A further set of columns were prone to producing false positives, but with a spread in time. To remove these, another iterator parsed the same $0.001°$-wide regions. Regions were masked if: (1) they contained more than ten events, (2) if the standard deviation along the row axis exceeded $0.01°$ and (3) fewer than 40 per cent of the events fell within $0.01°$ of the median position. This retains multiple flaggings of the same event, while masking large numbers of events occurring in separate locations.

This reduced the peak list to 9976 events at 8306 sky positions. The largely uniform sky distribution of these peaks can be seen in the upper panel of Figure 7.

Following this noise rejection process, we retain only three of the expected five OGLE events (OGLE-2016-BLG-1043, -1231, and -1245) and eight of the expected 21 KMTnet events (KMT-2016-BLG-0090, -0117, -0124, -0128, -0143, -0162, -0181, and -0189). Consequently, roughly half of the events visible in the lightcurves do not pass our detection pipeline.

## 6 LIGHTCURVE RE-EXTRACTION AND FITTING

### 6.1 Re-extraction

The MCPM extraction is model dependent (Section 2.2). We *a priori* assumed a flat lightcurve for each object. However, this small set of lightcurves can now be re-extracted, fitting a single-lens, single-source microlensing model. Accurate fitting is still impossible, as structured noise in the lightcurves (primarily from spacecraft motion and remaining variable stars) correlates strongly with the fitted event model. Nevertheless, a set of initial model fits can estimate the event properties, allowing further false positives to be removed. Note that we only use this re-extracted set of lightcurves in this section for the purposes of false-positive removal: we do not fully fit model parameters in this work.

Each peak was fitted using MULENSMODEL[13] (Poleski & Yee 2019), which interfaces with MCPM: we minimised the root-mean-squared residual between the model and lightcurve to both fit the lightcurve and improve the quality of the extracted photometry. We used the Nelder–Mead (downhill simplex) method of minimisation. The parameters $t_0$, $u_0$, $t_E$ and $F_{s,K2}$ (flux of the source star) were solved simultaneously. The first vertex of the simplex was initialised to the event time found by the peak finder, $u_0 = 1$ $R_E$, $t_E = 2$ days and $F_{s,K2} = 100$ e⁻ s⁻¹. The subsequent four vertices were respectively adjusted to three days before the peak-finder time, $u_0 = 0$, $t_E = 0.2$ days and $F_{s,K2} = 1000 \, e^- \, s^{-1}$. The MCPM minimiser tolerance was set to 0.0001. Only the sub-campaign (C9a or C9b) in which the event occurred was re-extracted. The bottom panel of Figure 7 shows the $u_0$ and $t_E$ extracted for each event. We remind the reader of the strong degeneracies between these parameters, particularly $F_{s,K2}$ and $t_E$. Consequently, these values are not to be taken as direct parametric measures, but merely as an aid to candidate selection in the next section.

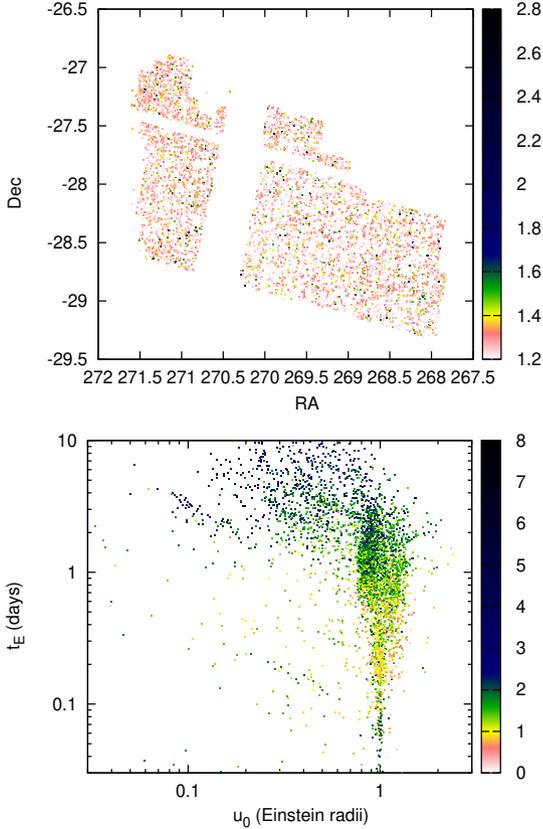

**Figure 7.** Top panel: sky locations of filtered *K2*C9 lightcurves. Colours are as in Figure 1. Bottom panel: fitted impact parameter and Einstein-crossing timescale for the same set of lightcurves, assuming them to be due to microlensing. The colour scale denotes the peak signal-to-noise (Section 6.1).

Many models were poorly fit, normally because the extracted signal did not resemble a microlensing event. These events result in the vertical stripe in Figure 7 around $u_0 = 1$ $R_E$. Real events are typically represented by many neighbouring spatial gridpoints, and scatter in diagonal lines towards lower $u_0$, highlighting the degeneracy between extracted $t_E$ and $u_0$ in the fitting procedure.

### 6.2 Noise filtering

The improved photometry and the fitted model allow better separation of false positives using the event's signal-to-noise ratio. To re-evaluate the lightcurve noise, we first perform the same two-day Gaussian filtering as previously, retaining both the low- and high-pass filtered components to (respectively) represent the temporally correlated ('red') and uncorrelated ('white') noise in the lightcurve.

Epochs outside the range $t_0 \pm 2t_E$ were used[14] to generate out-of-event noise amplitude for each component, determined as half the difference between the $5^{th}$ and $95^{th}$ centile fluxes over these out-of-event epochs. A median ($50^{th}$ centile) out-of-event flux was also established. The amplitude of the model, and the peak-to-median difference in observed flux were (respectively) used to derive two signal amplitudes: one for the model and one for the observed data.

A good candidate should have realistic fitted parameters and a peak substantially exceeding both the lightcurve's correlated and uncorrelated noise. Consequently, we remove peaks that do not fit all of the following criteria.

- $u_0 < 5R_E$. Peaks with larger $u_0$ will not produce significant magnification and are likely to be poorly fit models. This removes nine peaks.
- $t_E < 10$ days. Long-timescale peaks will not be FFPs. This cut corrects an issue from the above noise calculation, when no epochs exist to estimate noise from when $t_E$ is large. This removes 353 peaks.
- $t_E > 0.06$ days. Extremely short-timescale peaks are indistinguishable from artefacts or other short transients, and the existence of finite-source effects in true FFPs (i.e., it takes a finite time for the Einstein ring of the lens to cross the source's stellar disc) mean extremely short events are unlikely to have an origin in an FFP lensing event (cf. Mróz et al. 2020). This removes 300 peaks.
- A ratio between the observed signal strength (time-integrated flux within $t_0 \pm 2t_E$) and the red noise that exceeds 5. This removes 506 peaks.
- A ratio between the model signal strength and the red noise that exceeds 5. This removes 4443 peaks.
- A ratio between the observed signal strength and the white noise that exceeds 5. This removes 1256 peaks.
- More than three photometric datapoints within 2 $t_E$ of the fitted event centre that are above the 99th centile in both the original (flat-prior) and re-extracted (fitted-lens-prior) lightcurve. This removes 149 peaks.

Our peaks thus reduce from 9976 to 2960, occurring at 2229 sky positions. Single-linkage clustering suggests these represent ∼1770 discrete events. All remaining known events passed this cut. Visual inspection of these peaks shows that, while some data artefacts are still present, most remaining peaks appear astrophysical in nature.

## 7 DETAILED CANDIDATE SELECTION

In this Section, we continue to use automated cuts to remove false positives. However, each rejected peak is first subject to individual visual scrutiny. Peaks in lightcurves including suspected microlensing events are only rejected if other detections of the same event occur at the same sky position.

### 7.1 Variable star filters

Many remaining peaks are pulsating or ellipsoidal variable stars. In the flat-prior lightcurve, the complex PRF changes the amplitude of variability as the spacecraft moves, making some peaks stronger than others. MULENSMODEL then interprets strong peaks as microlensing events and, if the modelled blending fraction is high, an unreasonably high amplitude will be returned. A set of sequential filters remove these variables.

---

[14] Note the caveats in Section 6.1 regarding use of $t_E$.

Such stars have lightcurves that are largely symmetrical around zero flux. Events were rejected if the lowest point on the original (flat-prior) lightcurve (Section 2) is >7.5 times lower than the 16th centile of the lightcurve. Events were also rejected by comparing the faintest 0.2 centile (2nd permille) against the brightest (998th permille): if the second permille is further below the mean than the 998th permille is above the mean, the lightcurve is characterised better by dips than brightenings, triggering rejection. These cuts were quite effective, leaving 2462 peaks at 1822 sky positions.

The dominant frequency of each lightcurve was identified using a Lomb-Scargle periodogram of the flat-prior lightcurve, over the period range 0.1–10 days. A Ricker wavelet at this frequency was used as a matched filter on the lightcurve. Peaks were kept if they met both the following criteria: (1) that the ratio of the maximum of the cross-correlation function (CCF) within $t_0 \pm t_E$ to the maximum outside this range was $< \sqrt{N_p}$, where $N_p$ is the number of period cycles over the entire lightcurve (thus filtering strongly repeating variables); and (2), that the dominant period is less than half the fitted $t_E$ (thus retaining short, spike-like events).

Our final filter uses the amplitude of the CCF as a smoothed lightcurve. When subtracted from the flat-prior lightcurve, this creates a high-pass filtered lightcurve. The ratio of the absolute sum of the residuals of this filtered lightcurve to that the original lightcurve effectively provides the mean amplitude of the residual lightcurve after subtracting power at the strongest frequency. Peaks were rejected if the ratio in the sub-campaign during which the peak signal *did not* occur was <0.7, i.e., if a single sinusoid represents >30 per cent of the differential lightcurve. These cuts were effective at removing many pulsating stars, including many that appear to be rotationally modulated flare stars. The cuts leave 1994 peaks at 1433 sky positions.

## 7.2 Flare filters

Noise spikes now dominate the false positives, most being stellar flares on late-type dwarf stars, often identifiable as rotationally modulated variables with one or more sudden spikes in flux. These can be difficult to separate from very-short-timescale microlensing events. However, while microlensing events are symmetrical, stellar flares are skewed, brightening suddenly, then fading over a few exposures.

We fitted a skewed Gaussian to each peak with the SCIPY statistics package SKEWNORM via $\chi^2$ minimisation. A central timescale ($t_{0,skew}$), amplitude, width ($\sigma_{skew}$) and skew parameter ($a_{skew}$) were fit. Sources were rejected if any of the following criteria were met:

- $a_{skew} > 4$, provided $t_{0,skew}$ was within $\sigma_{skew} + 0.2$ days of the initially flagged event (to avoid fitting other events that are already rejected);
- $\sigma_{skew} < 0.02$ days, as these are universally single-epoch noise spikes (e.g., associated with cosmic ray hits);
- $\sigma_{skew} > 2$ days: FFP events should be shorter, and all such events exhibit sinusoidal variability;

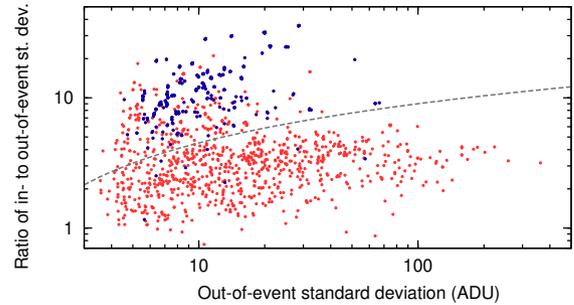

**Figure 8.** The relative signal-to-noise in a window around the peak, compared to the remainder of the lightcurve (Section 7.3.1). Darker, blue points mark known events; lighter, red points mark surviving peaks. Peaks below the dashed line are rejected. Known events below the line occur at different times from the event itself.

- sources with negative amplitudes, as these are also associated with variable stars;
- events peaking outside the observing window, or within one day of the start or end of a sub-campaign, and all such events are simply long-term photometric trends.

These cuts leave 1231 peaks at 898 sky positions.

## 7.3 More variable star filters

### 7.3.1 Peak significance filtering

Many peaks are still due to variable stars or sporadic noise spikes. Most have not yet been removed because the fitting functions have improperly characterised the event timescale. This is often because of multi-mode pulsations or correlated noise related to spacecraft motion.

To better characterise the timescale, a fixed $t_0$ was assumed, and a window of ±0.25 days selected around the peak. The average absolute deviation in the event ($\Gamma_{in}$) was calculated, and divided by the average absolute deviation in the remainder of the lightcurve ($\Gamma_{out}$). The windowing function was increased in increments of 0.01 days, up to 2 days, and the timescale with the maximum ratio of $\Gamma_{in}/\Gamma_{out}$ was chosen. This function is robust against events of different shapes.

Figure 8 shows this maximised ratio as a function of the corresponding $\Gamma_{out}$. The dotted line largely separates potential candidates from false positives, and is derived as

$$\Gamma_{in}/\Gamma_{out} = 4.5 \log_{10} \Gamma_{out}. \tag{12}$$

The increase in this ratio with $\Gamma_{out}$ is motivated by the tendency for lightcurves with large $\Gamma_{out}$ to be more greatly affected by correlated instrumental noise or pulsations of variable amplitude, rather than by random processes.

This cut is both efficient and severe (removing 718 peaks, leaving 513), hence it is performed at this late stage to allow visual confirmation: known false negatives (events spatially near microlensing sources) display a markedly different $t_0$ to the known event, and often very differently shaped lightcurves. Visual inspection also highlighted five rejected events that passed visual inspection for microlensing events. Of these, two ($\alpha = 268.473$, $\delta = -28.914$, $t_0 = 7516.93$ and $\alpha = 268.183$, $\delta = -28.782$, $t_0 = 7519.67$) are visually noticeable but comparatively faint; another ($\alpha = 268.090$, $\delta = -27.341$, $t_0 = 7522.93$) appears good, but the lightcurve suffers from correlated noise, modulated on a similar timescale to the event. The final two events are discussed in the next section.

Several of the remaining events are of poor-quality, but are retained as a single exposure within the event boosts their statistical weight. Therefore, a five-point median filter was applied to each lightcurve and the process run again, again removing objects failing the cut-off in Equation 12, or with $\Gamma_{in}/\Gamma_{out} < 5$. This removed an additional 25 peaks.

### 7.3.2 Additional asteroid filtering

The two remaining events discussed above were flagged at slightly different times. They occur when two asteroids pass over almost the same place at the almost same time, causing a single brightening of that spatial location. A single event was assigned, but offset in time from the ephemeris of either asteroid, hence missed by our earlier cut. Hence, we now remove any event within $0°002$ and 0.5 days of an asteroid passage. This removed 13 peaks, leaving 475 peaks, of which 277 are $<0°005$ and $<2$ days from existing OGLE or KMT-Net events. The other 198 peaks map to 97 sky positions, which clustering analysis associates with ∼66 unique events.

## 8 DETAILED EVENT SELECTION

### 8.1 Merging peaks

Multiple detections for each event arise from slightly different peaks identified in the first stage of the reduction. By now, only one physical brightening is associated with each lightcurve, merely flagged on multiple occasions. Hence, we can merge peaks in the same lightcurve together.

We first choose the peak with greatest signal in each lightcurve. This choice is not greatly important, as it only seeds the data reduction. We refit a skewed Gaussian to each peak using the SCIPY.STATS.SKEWNORM tool, but refining our previous technique as below. A Gaussian is used instead of a Paczynski function both to aid computation and because of its generalised application to non-microlensing peaks. The skew parameter is employed to fit stellar flares and outbursts.

Correlated noise (Section 4.4.2) means that a parametric fit contains many local minima, making it hard to find an accurate, global-minimum best fit. Consequently, we start with an initial Monte-Carlo fit to the lightcurve, with initial parameters of the peak time (Section 3.3), timescale (Section 7.3.1) and amplitude (Section 6.2), and zero skew. Over 10 000 Monte-Carlo iterations, a random perturbation in the event time was chosen from a normal distribution of width 0.5 days; the event width and amplitude were each multiplied by a factor drawn from a normal distribution with a unity mean and a standard deviation of 0.25; and the skew was drawn randomly from within the range −4 to +4. The fit with the lowest average squared deviation within two days of the supplied event time was then used as a seed for the numpy.optimize.minimize Powell minimisation function, which was run with default parameters.

This process was repeated for each peak, and the peak with the lowest average squared deviation after minimisation was selected to represent the best detection in that lightcurve, leaving lightcurves representing 215 spatial positions.

### 8.2 Symmetry filtering

The primary contaminants remain flares, and both instrumental and real step jumps in flux. The fitted Gaussian functions still have difficulty finding a global-minimum best fit, so were fitted again with 100 000 Monte-Carlo iterations. This improves the fits in many cases, though some lightcurves are still very poorly fit.

Events were removed as stellar flares if: (1) $|a_{skew}| > 5$ (most known microlensing events have $|a_{skew}| < 2$); and (2) the fitted Gaussian better fit the overall lightcurve than an unchanging baseline flux; and (2) the average squared deviation from the fitted Gaussian at timescales $>2$ days from the event increase by no more than 1 per cent compared to the baseline flux. This ensures that only well-fit, flare-like events were removed. This leaves lightcurves representing 153 sky positions, of which 85 are associated with known events.

This cut removed two known lenses that do not have Gaussian-like lightcurves: the binary event OGLE-2016-BLG-1043 and KMT-2016-BLG-0209. The latter event peaks ∼1 day from the start of the observations, and its beginning is not fully observed. The one-day training set used for photometric extraction at the start of the lightcurve is therefore not flat, and the event is not recovered symmetrically.

Visually inspecting the rejected lightcurves identifies two novel events that could be microlensing events, but are more likely correlated noise and a stellar flare. These occur respectively at $\alpha = 270.778$, $\delta = −27.511$, JD $= 2\,457\,505.34$ and $\alpha = 271.137$, $\delta = −26.920$, JD $= 2\,457\,551.64$.

This cut also affects binary microlensing lightcurves. Two out of three events were removed for an event at $\alpha = 269.880$, $\delta = −27.608$, JD $= 2\,457\,518.89$. The remaining event is discussed later. No other such events were visually detected.

### 8.3 Lightcurve merging

With a near-final set of lightcurves, we can now perform single-linkage clustering (as in Section 4.2) to merge associated lightcurves at neighbouring positions into individual events. Physical events are now cleanly separated from each other in space and time, allowing easy clustering.

For each event, the position with the largest-amplitude time-integrated signal was chosen as an initial localisation to represent the lightcurve. These positions undergo later refinement, so exact positions at this stage are not critical.

This produced 35 candidate microlensing events recovered from the K2C9 dataset with $t_{FWHM}$ consistent with an FFP, plus 22 associated with known lenses (after removing a cataclysmic variable; note we now include KMT-2016-BLG-2554 and -2583 in this list), inspected in Section 8.4. Our list of 35 candidate events are examined in Section 9.

### 8.4 Properties of known events

Full fitting of known events, combining ground-based data with K2, is deferred to future papers. Here, we use MCPM on these events a final time to formally fit $t_0$ and $t_{eff}$ directly from the K2C9 lightcurves. These fits are discussed in the Appendix.

**Table 3.** Events manually rejected as their lightcurves do not resemble microlensing events. Dates are JD − 2450000.

| RA | Dec | Date | Notes |
|---|---|---|---|
| 268.256 | −28.417 | 7521.134 | Flare |
| 268.523 | −28.308 | 7564.411 | Instrumental |
| 269.386 | −28.716 | 7542.186 | Periodic |
| 269.619 | −28.250 | 7555.556 | Flare |
| 269.643 | −28.869 | 7534.066 | Instrumental |
| 269.653 | −28.627 | 7561.789 | Instrumental |
| 269.671 | −27.480 | 7517.815 | Periodic |
| 269.871 | −27.653 | 7520.603 | Long |
| 270.001 | −27.562 | 7544.414 | Periodic |
| 270.160 | −28.452 | 7531.917 | Noise |
| 270.552 | −27.397 | 7515.642 | Flare |
| 270.802 | −28.391 | 7536.428 | Long |
| 270.951 | −27.220 | 7533.143 | Flare |
| 271.068 | −28.016 | 7535.643 | Long |
| 271.231 | −27.206 | 7566.090 | Flare |
| 271.362 | −27.059 | 7569.184 | Long |

**Table 4.** Events manually rejected as they do not resemble microlensing events after recentring. Dates are JD − 2450000.

| RA | Dec | Date | Notes |
|---|---|---|---|
| 268.1453 | −28.3631 | 7570.3924 | Instrumental |
| 268.2190 | −28.8010 | 7567.1473 | Variable/flares |
| 268.9946 | −28.5364 | 7525.1526 | Flare |
| 269.0820 | −28.7943 | 7559.0575 | Asteroid |
| 269.1930 | −28.2440 | 7551.2838 | Flare |
| 269.5045 | −28.2915 | 7514.5276 | Variable |
| 269.7970 | −27.5460 | 7543.8977 | Flare |
| 270.7014 | −28.1039 | 7547.4999 | Variable |
| 270.8377 | −27.3961 | 7541.6519 | Flare |
| 271.2873 | −27.1084 | 7536.7764 | Slow |
| 271.3748 | −27.2682 | 7564.3321 | Noise |

## 9 INSPECTION OF CANDIDATE EVENTS

### 9.1 Visual inspection

We now visually analyse the 35 objects selected for detailed inspection: the 16 listed in Table 3 are clearly associated with sporadic noise, poorly fit stellar flares, periodic variable stars, and instrumental artefacts. Most were not previously rejected because they were poorly fit by the Gaussian fitting process due to correlated noise in their lightcurves.

### 9.2 Object recentring

In this crowded field, the parameters used for the photometric extraction affect the final lightcurves, due to blending from nearby stars. Some candidates are both cotemporal with and within the PSF-radius of much stronger events, many of which have already been rejected. This was intentional: events must be extracted from within the PSF of variable stars, etc., so we must rely on lightcurves away from the PSF centre to detect subtle changes. However, we now need to recentre and re-extract these events to determine their true nature.

The flat-prior lightcurves within 0°003 of the candidate were windowed to the event timescale (i.e., the Gaussian width fitted in Section 8.2). The flux-weighted means of RA, Dec and JD of these lightcurves recentre the event.

The surrounding lightcurves were then inspected to ensure that only one event had been selected. In a few cases, this did not work correctly. One candidate ($\alpha = 268°145$, $\delta = −28°363$) lies on a chip boundary, while two other candidates ($\alpha = 269°716$, $\delta = −27°654$; $\alpha = 269°797$, $\delta = −27°546$) were each not correctly separated from a variable star. These positions were not updated. Another candidate ($\alpha = 268°222$, $\delta = −28°798$) was successfully relocated 0°004 away.

### 9.3 Re-evaluation of recentred lightcurves

Of the 19 remaining lightcurves, 11 have been recentred on non-microlensing events. Six of these events were previously rejected based on higher-quality lightcurves from the centres of their PSFs. Five events were rejected because structure before or after the event provides evidence against a microlensing event. Table 4 lists these rejects.

In four of the six in-event cases, the increased signal-to-noise ratio clarified the event as a stellar flare. A fifth event occurs near a chip edge, and is noise associated with the unprogrammed roll and correction. The final event was generated by asteroid 2009 BQ168. While our analysis was ongoing, the brightness estimated by JPL Horizons slightly increased: it is now expected to have brightened to marginally over 21st magnitude. This would barely warrant its inclusion in our asteroid list, but it would be undetectable. However, the recovered event is closer to magnitude 18.5 in the *Kepler* bandpass. Possible explanations include the magnitude of 2009 BQ168 being under-reported (the Horizons magnitudes are knowingly approximate); that the asteroid is exceptionally red, hence has a $(V − K_p)$ colour of ∼2.4 mag; or that the asteroid may have undergone an outburst.

Three of the five out-of-event cases show clear periodicity; one shows further stellar flares in its lightcurve. The 'slow' case (Table 4) shows a steep rise and slow decline, reminiscent of a stellar outburst. The 'noise' case showed periodic variability, neither clearly astrophysical or instrumental in nature. This leaves eight candidate novel microlensing events.

### 9.4 Object re-extraction and centring

The eight new candidates were then re-extracted using MCPM in a finer grid ($21''6 \times 21''6$, step size $0''72$), centred on their repositioned RA and Dec. The parameters were changed to extract only a smaller area, minimising effects of blended stars: four pixels were selected, and a training set of 100 pixels was used, with the ten largest outliers removed. Each lightcurve was also fitted with a microlensing model, using a tolerance ten times more strict than previously.

Using this grid, objects were re-extracted as previously. Events at $(\alpha, \delta) = (268.6613, −28.6140)$ and $(268.8945, −28.4774)$ were not recentred, multiple sources of variability contribute to their lightcurves. Table 5 lists the final positions, which we expect to be accurate to $\lesssim 2''$.

**Table 5.** Fitted parameters of novel candidate microlensing systems undergoing detailed inspection.

| K2C9- | *K2C9* Location | | Fitted parameters | | | Final |
|---|---|---|---|---|---|---|
| 2016- | RA | Dec | $t_0$ | $t_{eff}$ | $F_s/u_0$ | class |
| BLG- | (J2000) | | (JD)[1] | (days) | (e$^-$ s$^{-1}$) | |
| 1 | 267.9361 | −28.9215 | $7511.4562^{+0.0040}_{-0.0041}$ | $0.0535^{+0.0037}_{-0.0034}$ | $119.5^{+4.8}_{-4.7}$ | Passed, confirmed as astrophysical |
| 2 | 268.6613 | −28.6140 | $7544.5050^{+0.0009}_{-0.0009}$ | $0.0168^{+0.0009}_{-0.0008}$ | $313.5^{+10.0}_{-9.6}$ | Passed, unconfirmed |
| 3 | 269.7156 | −27.6540 | $7522.9878^{+0.0059}_{-0.0061}$ | $0.0616^{+0.0083}_{-0.0072}$ | $75.0^{+5.5}_{-5.2}$ | Passed, unconfirmed |
| 4 | 269.8359 | −27.5522 | $7558.0645^{+0.0036}_{-0.0035}$ | $0.0360^{+0.0042}_{-0.0037}$ | $109.3^{+7.3}_{-7.0}$ | Passed, unconfirmed |
| 5 | 269.8798 | −27.6074 | $7518.1435^3$ | —[2] | —[2] | Passed, confirmed as lens |
| — | 270.9172 | −27.2671 | $7561.0904^3$ | $0.852^3$ | $14.9^3$ | Rejected as noise |
| — | 271.2463 | −27.0928 | $7563.0933^3$ | $0.270^3$ | $12.8^3$ | Rejected as flare |
| — | 271.3306 | −27.2857 | $7535.4776^3$ | $1.369^3$ | $26.6^3$ | Rejected as variable |

[1] Barycentric Julian date - 2 450 000. Values for rejected lenses are at maximum flux. [2]Analysed in full in companion paper (Specht et al.), as a star–planet binary lens with ground-based detection. [3]Approximate values based on automatic modelling, assuming $u_0 = 1$.

### 9.5 Detailed inspection

Each candidate was checked against the Simbad[15] and VizieR[16] databases. Objects within one *Kepler* pixel were identified as potential counterparts, and objects $\lesssim 0\overset{\circ}{.}004$ away were considered contaminants. A summary of each is given below: three candidates are rejected and five remain.

#### 9.5.1 Rejected events

A — $\alpha = 270\overset{\circ}{.}9172$, $\delta = -27\overset{\circ}{.}2671$. The second recentering decreased significance and increased correlated noise related to spacecraft motion. Two nearby areas of strongly negative differential flux nearby could indicate hot pixels or flatfielding errors. The event is close ($0\overset{\prime\prime}{.}4$) to the $19^{th}$ magnitude star *Gaia* DR2 4063147780540495872, and lies in a very dense stellar field. As there is no clear event and clear correlated noise, we dismiss this candidate as instrumental noise.

B — $\alpha = 271\overset{\circ}{.}2463$, $\delta = -27\overset{\circ}{.}0928$. This is close ($0\overset{\prime\prime}{.}5$) to a $19^{th}$ magnitude star in the Panoramic Survey Telescope and Rapid Response System Data Release 1 (Flewelling et al. 2016), Pan-STARRS DR1 75482712462919121. The recentred lightcurve shows a cleanly extracted, symmetrical peak. However, the flux before and after the peak shows distinct asymmetry, so we reject this candidate as a stellar flare.

C — $\alpha = 271\overset{\circ}{.}3306$, $\delta = -27\overset{\circ}{.}2857$. This event is contaminated by a surrounding group of highly blended, $\sim 16^{th}$ magnitude stars. The recentred lightcurve shows a faint signal with a period of $\sim 5.6$ days, and it is likely that this unknown variable contributes a variable portion of the pixel's flux, as the other stars in the asterism move in and out of the pixel and contribute differently through the intra-pixel response. Consequently, we dismiss this candidate as a variable star.

#### 9.5.2 Accepted events

The final five events are assigned numbers (K2C9-2016-BLG-1 through -5), and their lightcurves shown in Figure 9. These represent data extracted by mcpm on the recentred candidate position, assuming a flat-prior lightcurve. Fitting of these events is performed in Section 11.1.2.

1 — This event is very well-isolated from contaminating variables and other stars, and cleanly extracted at $\alpha = 267\overset{\circ}{.}9363$, $\delta = -28\overset{\circ}{.}9216$. Only two objects exist within 4″: Pan-STARRS DR1 73292679366134218 ($1\overset{\prime\prime}{.}0$ away) 73292679356504208 ($1\overset{\prime\prime}{.}9$ away).

2 — This event is well-isolated but is centred near a pixel boundary. Consequently, the event is subject to substantial intra-pixel variation, so does not appear entirely symmetrical. Two DECaPS sources are recorded within 4″.

3 — This faint, but clean event has two components: one broad and one narrow. The narrow peak occurs a few hours after the broader one. One explanation is a caustic crossing of a binary lens, but this may still be an unusual stellar flare. The eclipsing binary OGLE BLG-ECL-224994 ($P = 1.28$ days) is 8″ away: this slightly contaminates the PSF, but does not significantly affect the lightcurve. Mild lightcurve structure is seen in the days following the event, however we accept this candidate on the balance of evidence.

4 — This source is close ($0\overset{\prime\prime}{.}5$) to the $18^{th}$ magnitude star *Gaia* DR2 4062799265608424320, and is $8\overset{\prime\prime}{.}9$ south of the bright ($B = 13.82$ mag) early-type star OPS 1759204-273316 (Grosbøl 2016). Noise from the latter causes the strongly positive and negative flux in Figure 10. The small-amplitude red giant OGLE BLG-LPV-160527 is 18″ away (Soszyński et al. 2013), but does not affect the lightcurve.

5 — This event is strong and cleanly extracted at $\alpha = 269\overset{\circ}{.}8799$, $\delta = -27\overset{\circ}{.}6079$. The lightcurve showed two peaks, typical of a binary microlensing event. The eclipsing binary OGLE BLG-ECL-232395 exists at 16″ from the target (Soszyński et al. 2016), but does not affect the lightcurve.

## 10 COMPARISON WITH EXTERNAL DATASETS

Despite our careful vetting, external comparison is needed to verify our candidates as microlensing events.

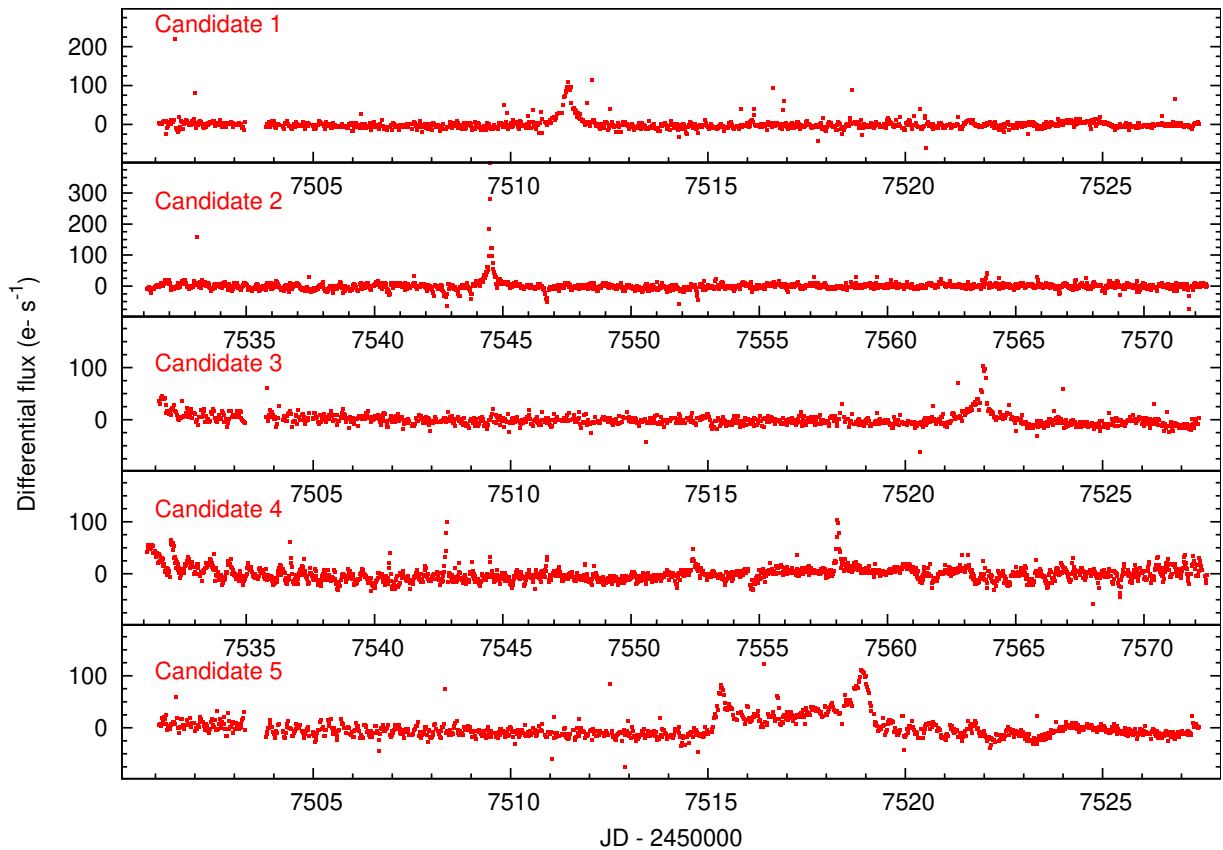

**Figure 9.** Lightcurves of new candidate microlensing events, extracted using MCPM by assuming a flat baseline flux model throughout, with identifiers from in Table 5. The data points are described in Section 9.

## 10.1 OGLE

Event 1 falls between a gap in the OGLE detectors, so was not recorded. Events 2, 3 and 4 were not clearly recorded due to gaps in the OGLE coverage caused by poor weather conditions. These gaps span the following periods:

2 — HJD 2457536.79 to 7545.59 (*K2* event at 7544.48);
3 — HJD 2457519.88 to 7522.65 (*K2* event at 7522.65);
4 — HJD 2457557.91 to 7561.51 (*K2* event at 7558.20).

Manual inspection of individual OGLE objects around these spatial locations could not find OGLE counterparts to our events. This suggests either: (a) that these events are not real or at much lower magnification at Earth; (b) that these events were sufficiently short-timescale not to show up in the OGLE data, despite OGLE observations often only narrowly missed the *K2* event peak; or (c) that these events were offset in time (by microlens parallax) so that they fall entirely within the gaps in the OGLE data. The example of KMT-2016-BLG-2554 shows that more than one of these reasons could be simultaneously true. Event 5 was recovered by OGLE at position $\alpha = 269.87983$, $\delta = -27.60747$ and shows multiple peaks between HJD 2457512 and 7517.

## 10.2 KMTNet

KMTNet data for these positions are not publicly available. Examination of nearby events suggests that all five candidates should have data within 24 hours of the *K2* peak, so should have been detected if they are true events that received similar magnification in the Earth's line of sight.

## 10.3 Canada–France–Hawaii Telescope (CFHT)

CFHT observed the *K2*C9 superstamp with the MegaCam wide-field imager several times per night in the *g-*, *r-*, and *i*-bands, aiming to measuring the colours of any short-timescale FFP candidates. MegaCam was only mounted on the telescope for four two-week runs at or around the time of the *K2*C9. Full details of the observations and their reduction are provided by Zang et al. (2018). Difference images produced using ISIS (Alard & Lupton 1998; Alard 2000) were visually inspected within ~6″ of the candidate locations for the duration of the observations.

Candidate 1 returns a strong visual source at $(\alpha, \delta)_{ICRS} = (267.93612, -28.92149)$ in two *i*-band difference images taken at BJD′ = 7511.02 and 7511.98, but not in the next two nearest images at BJD′ = 7510.12 and 7513.04. The source was not visible in *r-* or *g*-band images taken nearly coincidentally, but the region is heavily reddened ($A_i = 3.93 \pm 0.04$, $E(r-i) = 1.54 \pm 0.02$, estimated based on the location of the red clump using the same method as (Nataf et al. 2013) and assuming the intrinsic color of the clump from (Nataf et al. 2016)). Photometry at the location in the *r*-band reveal a marginal detection of flux in a single image at BJD′ = 7511.98.

Candidate 2 was observed in two overlapping CFHT fields, CF1 and CF6. Observations were obtained at least daily dur-ing the third MegaCam run ($\text{BJD}' = 7538.1$ and $7548.0$) with only one gap, between $\text{BJD}' = 7545.07$ and $7546.94$. No source was visually identified in CF6, but in CF1 a marginal visual detection was identified at $(\alpha, \delta)_{ICRS} \approx (268.66129, -28.61404)$ in an $i$-band image at $\text{BJD}' = 7542.91$. The nearest CF6 image was taken 6 minutes later, so we expect that the visual detection in CF1 was spurious. A possibility remains that a short event occurred in the one-night observing gap.

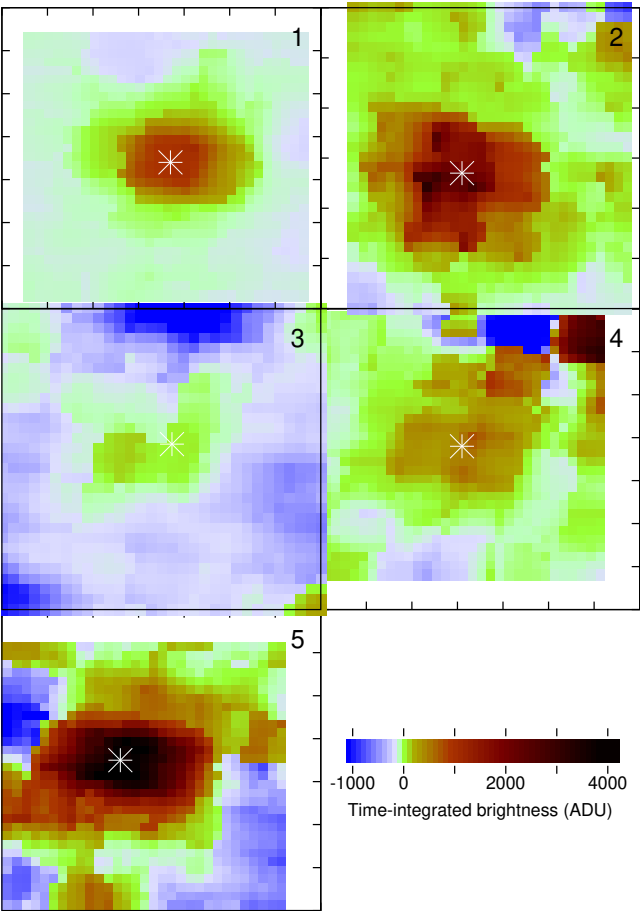

**Figure 10.** Relative integrated brightness of events over a sub-pixel-sampled region. Each box is $0°\!.006 \times 0°\!.006$. Brightness increases towards darker, redder colours, and is defined as the integrated flux of the extracted (differential) lightcurve within the time range $t_0 \pm \tau_0/2$. Negative brightness show decreased flux compared to the baseline. White asterisks mark the adopted location of each source.

Candidate 3 was not visually detected, but the $K2$ event was at the end of a MegaCam run. If the ground-based event peaked after the K2 event, we would not expect a detection.

Candidate 4's event occurred between MegaCam runs and was likewise not visually detected.

Candidate 5 was visually detected in several images in $g$-, $r$-, and $i$-bands at $(\alpha, \delta)_{ICRS} = (269.87984, -27.60736)$ between $\text{BJD}' \sim 7511$ and $7516$. The lightcurve shows two sharp peaks upon a shallow decline throughout the observations, indicative of a planetary or secondary caustic crossing in the wing of a microlensing event with a reasonably long timescale.

## 11 DISCUSSION

### 11.1 Candidate properties

#### 11.1.1 Summary of status

Our analysis has revealed five new candidate microlensing events in the $K2C9$ data. Candidates 3 and 4 have no ground-based observations. Comparative ground-based lightcurves for candidates 1, 2 and 5 are shown in Figure 11, and in more detail in the Appendix. For candidates 1 and 5, we update the astrometry (Table 5) based on the average of the OGLE and CFHT positions.

Visual inspection of the CFHT images suggests candidate 1 may be associated with Pan-STARRS DR1 73292679366134218, which is recorded at $i = 22.173 \pm 0.077$, $z = 20.820 \pm 0.099$, $y = 20.493 \pm 0.223$ mag in that survey. The strong drop-off with increasing wavelength is consistent with a cool or highly extincted star, explaining the significant differences in differential flux between the CFHT $r$- and $i$-band lightcurves. Ground-based data show a peak that is roughly co-incident with the $K2C9$ peak, showing this is an astrophysical event, consistent with a microlensing source. However, without a clear time or magnitude offset, we cannot rule out other kinds of event (e.g., stellar flares). Hence, this object remains a candidate event only.

Candidates 2, 3 and 4 show the hallmarks of real events but, without ground-based confirmation, we cannot correctly assign them to any particular category. Given the availability of terrestrial observations, we treat candidate 2 as a potential microlensing event with no perceptible magnification on Earth, but stop short of claiming it as a real event. We treat candidates 3 and 4 as suspected events, but without the ability to confirm or refute them. In particular, we note that candidates 3 and 4 have noticeable asymmetries in their lightcurves: these are small, and ultimately could be attributable to the structured noise terms seen in real events (see Appendix, e.g., OGLE-2016-BLG-014 and -1231, and KMT-2016-BLG-0162).

#### 11.1.2 Fitting of candidate parameters

In Poleski et al. (2019), the analyzed events had ground-based data that constrained $t_E$, which is usually the same to within a few percent for ground and satellite observations of a given event[17]. Our four new single-lens candidate events have very sparse ground-based data that do not allow the event timescale to be meaningfully constrained. To measure the microlensing parallax using the timescale measured in the $K2$ data, one needs to constrain $t_{0,\oplus}$ plus the source and blending fluxes from ground-based data. However, the existing data are too sparse to constrain these parameters. The sparcity of the ground-based data is easily understood: if these events had more epochs from microlensing surveys, it is likely that the surveys would have discovered them in the first place. Hence, we decided to only fit the $K2$ data without constraining the microlensing parallax.

---

[17] The difference in timescale of the event for different observers will depend on their relative velocity tangential to the line of sight to the event ($\lesssim 30$ km s$^{-1}$) and the projected velocity of the lens-source line of sight across the solar system ($\sim 100$-1000 km s$^{-1}$, e.g., Calchi Novati et al. 2015).

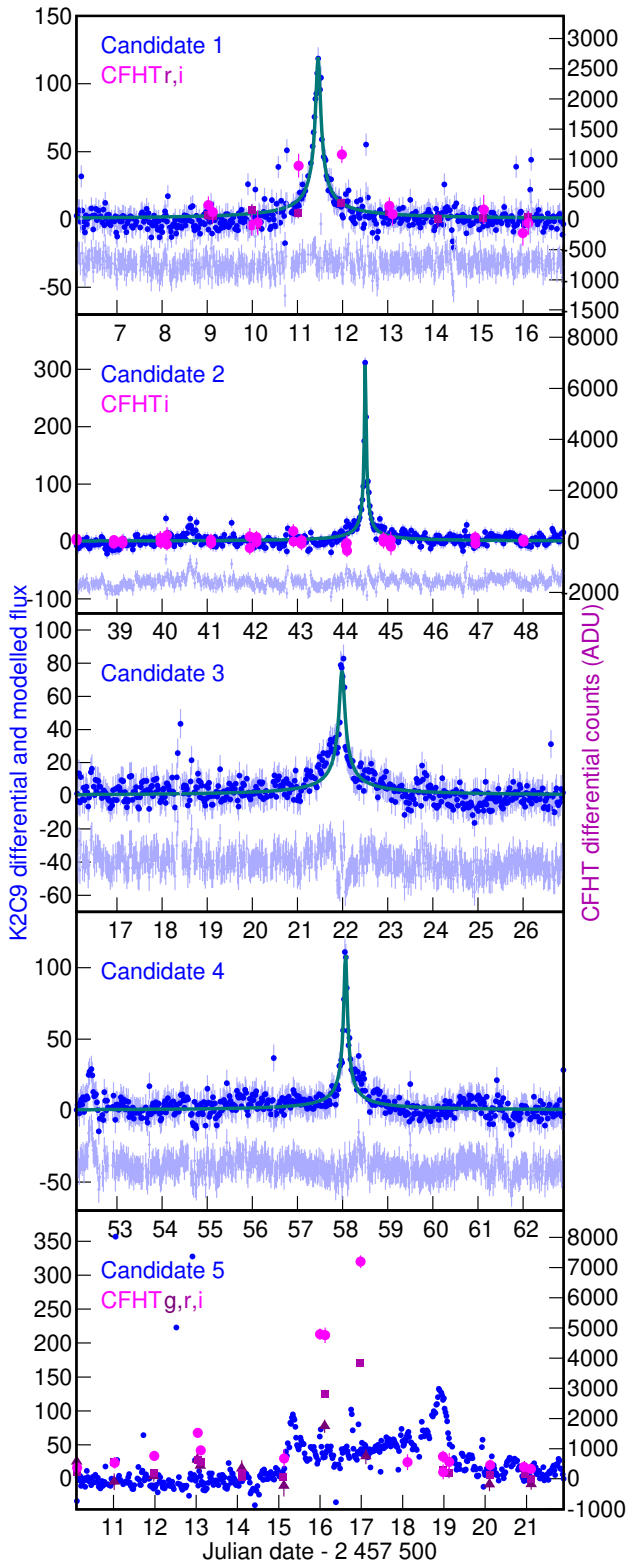

**Figure 11.** *K2*C9 lightcurves of new FFP candidates with ground-based observations. *K2* observations are in shown as blue points, with the corresponding model as a cyan line. Residuals are shown in light blue, offset below. Candidate 5 is modelled separately in Specht et al. (in prep). Observations from CFHT are shown in purple, approximately scaled to the same intrinsic flux.

Mcpm has not previously been used for events without ground-based data. To test the method, we fitted a four-parameter Paczyński model (with zero baseline flux 0) to the *K2* data for events analyzed by Poleski et al. (2019). These fits revealed that the model is too flexible and it overfits the noisy data, resulting in significantly underestimated $t_E$. The other possible approach is to use a pixel-lensing model, in which the excess flux is fitted by a three-parameter model: $t_0$, $\Delta F \equiv F_s/u_0$ (where $\Delta F$ is the change in flux and $F_s$ is the source flux), and $t_{eff}$ (Gould 1996). The smaller the value of $u_0$, the better the pixel-lensing approach approximates Paczyński equation. Other than this simplified model, the fitting followed Poleski et al. (2019), i.e., the lightcurve was simultaneously extracted and fitted, the photometry was extracted for four central pixels, 500 training pixels were used in the first step, then limited to 100 pixels, and L2-regularisation factor was set to 6000 per pixel. As a consistency check, we fitted events analyzed by Poleski et al. (2019) and found that the formal uncertainties on the fitted $\Delta F$ and $t_{eff}$ are underestimated, but that these parameters are accurate to $\approx 10\%$. The results of the parameter inference for the four new candidate events are presented in Table 5 and shown in Figure 11. The fits for the previously known OGLE/KMTNet events that were selected by our pipeline are presented in the Appendix. We note that for candidates 1 and 2, the peaks of the final lightcurves are brighter than those extracted in Section 2.2.

### 11.2 A new bound microlensing event

Candidate 5 is clearly recovered in terrestrial data, but with a lightcurve with peaks that do not match the timing from *K2*. These differences show that the event is not intrinsic to the source, but caused by a previously unknown microlensing event. The shape of the event in *K2* and ground-based data is consistent with a binary lens, with the primary-lens peak visible in only the ground-based data, near HJD 2457513. Caustic crossings then occur in the ground-based data at dates $\approx$ 7515.6 and 7516.9, and in the *K2* data at dates 7515.3 and 7518.9. Visual inspection of the CFHT images shows the event to be associated with a faint but resolved star, currently without a designation. It is unresolved from two close neighbours and two nearby brighter stars (Pan-STARRS DR1 74872698800561822 and 74872698803761694) in the PanSTARRS and DECaPS surveys.

Modelling of this lens (presented in a companion paper, Specht et al., in prep) identifies it as having a mass ratio consistent with a low-mass star with a Jovian-mass planetary companion. As of April 2021, there are 149 known microlensing planets[18], though this would be the first microlensing planet to be discovered from a space-based observatory (all previous events that have included space-based data were first identified or alerted from ground based observations).

### 11.3 Free-floating planets versus stellar-mass lenses

Although FFP microlensing events separate well from stellar-mass events in $t_E$, the same has not been demonstrated for $t_{eff}$. To show this, we have simulated a population of microlensing events using the MaBμLS-2 simulator ([Specht et al. 2020](#)) optimised to *K2* parameters. By computing both the $t_E$ and $t_{eff}$ for each simulated microlens, we can compute the stellar and brown dwarf $t_{eff}$ distribution shown in black in Figure 12.

---

[18] http://www.exoplanet.eu/

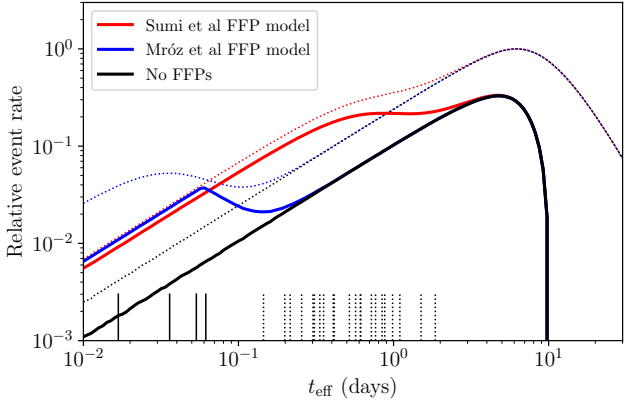

**Figure 12.** The predicted $t_{eff}$ distribution for microlensing events towards the *K2C9* field for stars and brown dwarfs (black), with the addition of either the Jupiter ([Sumi et al. 2011](#), in red) or Earth ([Mróz et al. 2017](#), in blue) FFP models. Thin, dotted lines show the full distributions, in the limit of uniform detection efficiency. Thick, solid lines show the distributions when events are windowed to $0.06 < t_E < 10$ days. Vertical lines at the plot's base show our four candidate FFP events (solid lines), and our $t_{eff}$ fits to *K2C9* data of the OGLE/KMTNet events presented in the Appendix (dashed lines).

For FFP calculations we replace each lensing star with a planetary-mass lens that inherits its position and kinematics, thus generating the expected $t_{eff}$ distribution for Jupiter and Earth mass FFP populations. Scaling these to 1.8 Jupiter-mass FFPs per star (as discussed in [Sumi et al. 2011](#)), and 10 Earth-mass FFPs per star (as discussed in [Mróz et al. 2017](#)), and adding them to the stellar-mass distribution, we generate the combined FFP–stellar $t_{eff}$ distributions shown as coloured lines on Figure 12.

These distributions assume an event recovery that is independent of $t_{eff}$. In reality, the recovery efficiency depends on the relative sensitivity of the underlying photometry, and the selection effects of our pipeline's smoothing functions. The efficiency function cannot be described without complex simulations ([Specht et al., in prep](#)), but we expect roughly equal sensitivity over the range $0.083 \lesssim t_{eff} \lesssim 2$ days, with a smooth decline in sensitivity at both ends. The solid lines in Figure 12 show the effect of a more conservative windowing function of $0.06 < t_{eff} < 10$ days that is applied in our pipeline.

Figure 12 also indicates the $t_{eff}$ of our four new candidate FFP events, and the previously known events from OGLE and KMTNet that pass our selection pipeline. The known events scatter around $t_{eff} \approx 0.6$ days, and are in line with the short-$t_{eff}$ tail of a stellar distribution. The lack of sources with $t_{eff} \gtrsim 1$ day implies declining sensitivity beyond this point. The $t_{eff}$ of our four candidate FFP sources is an order of magnitude or more shorter than this peak. While firm conclusions can only be drawn after we complete our efficiency analysis ([Specht et al, in prep](#)) their timescales are consistent with the peak expected for the Earth-mass FFP model. As such they are perhaps the most promising FFP candidates recovered from *K2*.

## 11.4 Frequency of free-floating planets

A thorough evaluation of the FFPs frequency is in progress ([Specht et al., in prep.](#)) and requires a Galactic model description of the FFP distribution and a determination of the event recovery efficiency through our *K2C9* selection pipeline. Nevertheless, we can here make some quantitative statements on the FFP abundance and the probability that our candidates represent a genuine FFP population.

[Penny et al. (2017)](#) calculated the potential *K2C9* FFP yield, estimating that 1.1–6.3 events would be detectable (based on the higher [Sumi et al. (2011)](#) FFP abundance, and including the 20 per cent loss due to *K2* entering emergency mode). The majority (roughly 60–85 per cent) of these would be simultaneously detectable by ground-based observatories, with most of the remainder having an Einstein ring that crosses *Kepler's* orbit but not Earth's. Crucially, short-timescale FFPs separate well from stellar lenses in terms of their Einstein radii (e.g., [Penny et al. 2017](#), their figure 3), with FFPs peaking around the separation of *K2* from Earth, and stars having much larger Einstein radii. This means that single-lens events (where caustic crossings do not exist) present large differences in amplitude between *K2* and Earth if the lens is an FFP, but not if the lens is a star. Candidate FFPs therefore include

   (i) *K2* events not recovered by terrestrial observatories;
   (ii) terrestrial events that are not recovered by *K2*;
   (iii) events detected at both sites with large time and/or magnification differences.

The first category includes our events K2C9-2016-BLG-2, -3, and -4, which appear to have been undetectable from Earth. If these are indeed astrophysical events, their lack of terrestrial detection adds weight to their being true microlensing events with projected Einstein radii $\lesssim 1$ AU. However, without ground-based confirmation, we cannot determine their nature with confidence. Consequently, we can say that there are 0–3 detected events in this category.

The second category contains some of the events in Section 3.1 that were not visible in the *K2* lightcurves: five OGLE events (OGLE-2016-BLG-1041, -1058, -1110, -1097, and -1162) and 10–18 KMTNet events, depending on whether the events weakly visible by eye are included. Generally, these events are confirmed as astrophysical via observation by more than one terrestrial observatory. However, the heavy blending and sensitivity differences (notably due to the bluer $K_p$ filter and heavy interstellar reddening of the source stars) mean we cannot expect all of these would be detectable by *K2*, but this would require extensive modelling to determine the dilution by blending. There are also significant uncertainties on the $t_E$ and/or $t_{eff}$ of some events, particularly in the KMTNet dataset, hence we cannot confidently determine the number of short-timescale microlensing events either.

The third category does not include any of our novel FFP events, but does include K2C9-2016-BLG-5 that is due to a bound planet. Some of the four events discussed in Section 8.4 may also fall into this category.

Whilst we cannot yet unambiguously confirm FFPs from the *K2C9* dataset, we have identified several short-timescale events that are promising candidates. We know from events such as OGLE-2016-BLG-1928 ([Mróz et al. 2020](#)), which lies just outside the *K2C9* field of coverage, and OGLE-2016-BLG-1540 (Mróz et al. 2018), which occurred inside the field of coverage but after the cessation of observations, that FFPs can be expected. Hence, at this point, it is certainly credible that the numbers of recovered FFPs are in line with Penny et al. (2017) estimates. Comparison against other (proprietary) sources of ground-based data may confirm some of our novel candidates, and improved modelling of our recovery of known ground-based candidates may confirm some of these short-timescale events as FFPs.

### 11.5 Lessons for future microlensing space missions

#### 11.5.1 Event detection and isolation

*K2C9* (and previous *Spitzer* surveys, e.g. (Calchi Novati et al. 2015; Yee et al. 2015)) act as a pathfinders for dedicated microlensing surveys using upcoming space-based facilities: a significant effort has already been expended formulating microlensing surveys for *Euclid* (Laureijs et al. 2011) and the *Nancy Grace Roman Space Telescope* (formerly *WFIRST* and hereafter refered to as *Roman*; Green et al. 2012; Penny et al. 2019; Johnson et al. 2020). This work demonstrates that blind space-based microlensing surveys are possible. Indeed the considerable challenges posed by the sub-optimal nature of the *K2C9* dataset for microlensing analyses provide an extreme stress-test that allows us to push forward into the space-based microlensing era with confidence.

The most significant issues affecting *K2C9* were ultimately due to (in order of precedence): the telescope's large (4″) pixel size compared to the high sky density of stars towards the Galactic bulge, the undersampling of the PSF with respect to this pixel size, and the spacecraft motion caused by the lack of gyroscopic control in the extended *K2* mission.

*Pixel scale: Kepler's* large pixel scale means each pixel represents the overlapping PSFs of many stars and asteroids: the Dark Energy Camera Plane Survey survey (DECaPS Schlafly et al. 2018), the deepest large-scale survey of the region, averages ∼7 stars per *Kepler* pixel, not including stars missing due to its own blending incompleteness, and the many microlensing sources below the detection threshold. This makes event localisation very difficult, but also means that issues affecting a small percentage of stars can make large areas of the detector unusable. In *K2C9*, this included variable stars, many of which have not been adequately catalogued, especially low-amplitude or sporadic variables. Saturated stars and asteroids also mean large regions of the detector are masked, but these can be more easily identified and accounted for. Pixel artefacts also posed significant problems, including those with large dark currents and/or are poorly flat-fielded.

Both *Euclid*'s visible instrument and *Roman*'s near-infrared camera will observe several magnitudes deeper than *K2C9*. However, their pixels (0″1 and 0″11, respectively) have areas ∼1500× smaller, more than compensating for the increased source density. Simulated images, using population-model data (Penny et al. 2013, 2019), suggest that individual stars will be much more cleanly resolved than with *K2* (cf. Penny et al. 2017), and that microlensing sources will be much more cleanly identified. Hence, we expect these to be much less severe for *Euclid* and *Roman*.

*PSF undersampling and aberration: Kepler* suffered from significant PSF undersampling: 20–62 per cent of a star's flux fell in a single pixel[19], and the strong intra-pixel response function varied by a factor of two across a pixel (Bryson et al. 2010). Without knowing an event's precise intra-pixel co-ordinates, these led to the significant variations in the flux received from both the source star and its blended neighbours, producing the strong correlated-noise signals of our data. This occurs despite the telescope PSF and camera intra-pixel response function being well-characterised. It also exacerbates small errors in dark subtraction and flat-fielding: these alone often dominate over the photonic Poisson noise.

Additionally, *Kepler* exhibited strong aberration. The *K2C9* super-stamp was at the edge of the field of view, due to ecliptic-plane restrictions. This led to a bright PSF core, causing the problems just highlighted, while spreading a small fraction of the star's flux out over many pixels, exacerbating the aforementioned problems of stellar crowding.

*Euclid* and *Roman* both have 50 per cent encircled energy radii of ∼0.13 arcsec (i.e., half the star's flux is contained within this radius, or within central ∼4–6 pixels)[20]. Consequently, the PSF undersampling is still significant, but a factor of several less than *Kepler*. The intra-pixel quantum efficiencies for *Euclid* VIS and *Roman* have yet to be formally released. Their smaller overall field of view and on-axis pointing will also reduce the effects of PSF aberration.

*Spacecraft motion:* broken reaction wheels meant *K2* lacked good gyroscopic control. *Euclid* and *Roman* will not have such problems. Diffraction spikes from bright stars, caused by the secondary mirror supports (three for *Euclid*, six for *Roman*) may rotate around the frame, causing spurious brightenings. However, most of the problem sources in *K2C9* were caused by blended stars, not the motion on the detector of the source star images themselves, so we correlated noise to be a present but lesser problem for the future missions.

#### 11.5.2 False-positive rejection

In the *K2C9* data, false positives were removed in approximate order of ease of separation from true microlensing signals. Most artefacts related to spacecraft motion can be removed through correlation of the lightcurve with the spacecraft pointing vector (though the correlation will be different for every star). We have shown that these motions can be accounted for relatively easily and, with a better-sampled PSF and better resolution, such false positives should be much easier to characterise. Similarly, with a wider PSF, a sharpness criterion could be used to differentiate cosmic-ray impacts from stellar brightening events.

Variable stars and asteroids are significant false positives in our data. Recognition and removal of asteroids was fairly straight-forward, but the variable star catalogues we used are incomplete. We can expect their completeness to improve through ongoing surveys like OGLE, and new surveys like the Vera Rubin Observatory (formerly LSST; Ivezić et al. 2019).

One of the last false-positive categories we removed was stellar flares. In the era of *Euclid* and *Roman*, these may prove the most tenacious category of false positive. With better resolution and PSF sampling comes lower instrumental noise, and it should be easier to determine whether lightcurves are genuinely skewed, rather than merely appearing so due to the varying PRF. However, the fundamental limit to detecting skewness is the number of exposures that happen during the event. Many of the stellar flares in this dataset are not strictly sudden onset, but occur after unresolved "foreshocks", and still lead to a markedly skewed lightcurve, but not an instantaneous brightening. In general, our 30-minute cadence produced a limit whereby events shorter than $\sim$2–4 hours could not readily be differentiated from stellar flares. However, we must bear in mind that this is partly due to the need for adaptive photometric methods like MCPM due to the large pixel scale. Analysis of *Euclid* and *Roman* data will make use of highly precise difference image photometry pipelines, allowing very high-precision cuts to be implemented on overall lightcurve smoothness and symmetry that will provide greater screening against stellar flare event.

---

[19] https://keplerscience.arc.nasa.gov/the-kepler-space-telescope.html.
[20] https://roman.gsfc.nasa.gov/science/sdt_public/WFIRST-AFTA_SDT_Report_150310_Final.pdf.

### 11.5.3 Event confirmation

Despite the limitations of the $K2$ mission, we have shown the viability of space-based microlensing surveys. We have also demonstrated their potential to identify and characterise parallax effects when combined with ground-based observations, leading to the confirmation of exoplanetary candidates.

This added dimensionality, coming from taking multiple chords through the Einstein ring of the microlens, is only possible with a good quality of ground-based observation. The new events described here were primarily missed from ground-based observatories because they did not coincide with either the hours of darkness or clear weather. Consequently, if future space-based microlensing surveys have complementary observations from ground-based observatories with high ($\lesssim$hourly) cadence observations and redundancy for weather and technical limitations, the events that are bright enough to be detected from the ground will be characterized better.

## 12 CONCLUSIONS

We have performed a blind search for new candidate microlensing events in the $K2$ Campaign 9 observations of a 3.7 $\text{deg}^2$ region of the Galactic bulge, focussing on detection of systems with Einstein-crossing timescales between approximately 2 hours and 2 days. Considering all timescales, we recovered 22 known events and detected four FFP candidates, and one new, bound planet. One of our candidates was recovered in the ground-based data, confirming it as astrophysical, but for none of these events has $t_{\rm E}$ been measured reliably. Thus, the number of FFP candidates from $K2$C9 is between zero and four. This number was predicted by Penny et al. (2017) to be 1.1–6.3 based on the Sumi et al. (2011) FFP rate. The FFP rate was measured by Mróz et al. (2017) using a larger sample of events and found an upper limit seven times smaller than the Sumi et al. (2011) result, which accordingly scales the Penny et al. (2017) prediction.

In this analysis, a bespoke reduction process was necessary due to the poor resolution and complex systematics of $K2$. Looking ahead to upcoming space missions (*Euclid* and

*Roman*), we are optimistic they can recover short-timescale microlensing events, as the systematic issues that affected $K2$ should be less significant. Our most significant concern with these new missions is their ability to accurately separate short-timescale events from stellar flares and have sufficient ground-based follow-up so that short-timescale events can be reproduced despite daylight and weather restrictions.

## ACKNOWLEDGEMENTS

EK acknowledges funding by the UK Science and Technology Facilities Council (grant ST/P000649/1). This paper includes data collected by the K2 mission. Funding for the K2 mission is provided by the NASA Science Mission directorate. Some of the data presented in this paper were obtained from the Mikulski Archive for Space Telescopes (MAST). We thank OGLE team for sharing results of their observations. Work by RP was supported by Polish National Agency for Academic Exchange grant "Polish Returns 2019." This research uses data obtained through the Telescope Access Program (TAP), which has been funded by the National Astronomical Observatories of China, the Chinese Academy of Sciences (the Strategic Priority Research Program "The Emergence of Cosmological Structures" Grant No. XDB09000000), and the Special Fund for Astronomy from the Ministry of Finance. This work was performed in part under contract with the California Institute of Technology (Caltech)/Jet Propulsion Laboratory (JPL) funded by NASA through the Sagan Fellowship Program executed by the NASA Exoplanet Science Institute. Work by M.T.P. was partially supported by NASA grants NNX16AC62G and NNG16PJ32C, and Louisiana Board of Regents Support Fund (RCS Award Contract Number: LEQSF(2020-23)-RD-A-10. Based in part on observations obtained with MegaPrime/MegaCam, a joint project of CFHT and CEA/DAPNIA, at the Canada–France–Hawaii Telescope (CFHT) which is operated by the National Research Council (NRC) of Canada, the Institut National des Science de l'Univers of the Centre National de la Recherche Scientifique (CNRS) of France, and the University of Hawaii. The authors wish to recognize and acknowledge the very significant cultural role and reverence that the summit of Mauna Kea has always had within the indigenous Hawaiian community. We are most fortunate to have the opportunity to conduct observations from this mountain. S.M. and W. Zang acknowledge support by the National Science Foundation of China (Grant No. 11821303 and 11761131004). W. Zhu was supported by the Natural Sciences and Engineering Research Council of Canada (NSERC) under the funding reference #CITA 490888-16. We finally thank the anonymous referee for constructive feedback.

## DATA AVAILABILITY

The underlying data for this work is in the public domain, and sourced from the Mikulski Archive for Space Telescopes, https://archive.stsci.edu/k2/. Third-party data from OGLE and the CFHT were provided by co-authors RP and MP, respectively, and will be shared on reasonable request to the corresponding author, subject to the permission of the relevant observing groups. The analysis tools developed for this article are considered bespoke, but can be shared on reasonable request to the corresponding author.

# REFERENCES

Alard C., 2000, A&AS, 144, 363
Alard C., Lupton R. H., 1998, ApJ, 503, 325
Ban M., Kerins E., Robin A. C., 2016, A&A, 595, A53
Bihain G., et al., 2009, A&A, 506, 1169
Borucki W. J., et al., 2010, Science, 327, 977
Bryson S. T., et al., 2010, ApJ, 713, L97
Calchi Novati S., et al., 2015, ApJ, 804, 20
Clanton C., Gaudi B. S., 2017, ApJ, 834, 46
Di Stefano R., 2012, ApJS, 201, 20
Einstein A., 1936, Science, 84, 506
Flewelling H. A., et al., 2016, arXiv e-prints, p. arXiv:1612.05243
Gahm G. F., Grenman T., Fredriksson S., Kristen H., 2007, AJ, 133, 1795
Gaia Collaboration et al., 2018, A&A, 616, A1
Gould A., 1996, ApJ, 470, 201
Gould A., Horne K., 2013, ApJ, 779, L28
Green J., et al., 2012, arXiv e-prints, p. arXiv:1208.4012
Grošbøl P., 2016, A&A, 585, A141
Halbwachs J. L., Arenou F., Mayor M., Udry S., Queloz D., 2000, A&A, 355, 581
Han C., 2006, ApJ, 644, 1232
Han C., Chung S.-J., Kim D., Park B.-G., Ryu Y.-H., Kang S., Lee D. W., 2004, ApJ, 604, 372
Helstrom C. W., Wilson F. L., 1970, Physics Today, 23, 73
Henderson C. B., Shvartzvald Y., 2016, AJ, 152, 96
Henderson C. B., et al., 2016, PASP, 128, 124401
Howell S. B., et al., 2014, PASP, 126, 398
Ivezić Ž., et al., 2019, ApJ, 873, 111
Johnson S. A., Penny M., Gaudi B. S., Kerins E., Rattenbury N. J., Robin A. C., Calchi Novati S., Henderson C. B., 2020, AJ, 160, 123
Kim S.-L., et al., 2016, Journal of Korean Astronomical Society, 49, 37
Kim D. J., et al., 2018a, AJ, 155, 76
Kim H. W., et al., 2018b, AJ, 155, 186
Laureijs R., et al., 2011, preprint, (arXiv:1110.3193)
Libralato M., Bedin L. R., Nardiello D., Piotto G., 2016, MNRAS, 456, 1137
Lucas P. W., Roche P. F., 2000, MNRAS, 314, 858
Ma S., Mao S., Ida S., Zhu W., Lin D. N. C., 2016, MNRAS, 461, L107
Mróz P., et al., 2017, Nature, 548, 183
Mróz P., et al., 2018, AJ, 155, 121
Mróz P., et al., 2020, ApJ, 903, L11
Nataf D. M., et al., 2013, ApJ, 769, 88
Nataf D. M., et al., 2016, MNRAS, 456, 2692
Paczyński B., 1986, ApJ, 304, 1
Peña Ramírez K., Béjar V. J. S., Zapatero Osorio M. R., 2016, A&A, 586, A157
Penny M. T., et al., 2013, MNRAS, 434, 2
Penny M. T., Rattenbury N. J., Gaudi B. S., Kerins E., 2017, AJ, 153, 161
Penny M. T., Gaudi B. S., Kerins E., Rattenbury N. J., Mao S., Robin A. C., Calchi Novati S., 2019, ApJS, 241, 3
Poleski R., Yee J. C., 2019, Astronomy and Computing, 26, 35
Poleski R., Penny M., Gaudi B. S., Udalski A., Ranc C., Barentsen G., Gould A., 2019, A&A, 627, A54
Quanz S. P., Goldman B., Henning T., Brandner W., Burrows A., Hofstetter L. W., 2010, ApJ, 708, 770
Rasio F. A., Ford E. B., 1996, Science, 274, 954
Schlafly E. F., et al., 2018, ApJS, 234, 39

Schneider A. C., Windsor J., Cushing M. C., Kirkpatrick J. D., Wright E. L., 2016, ApJ, 822, L1
Soszyński I., et al., 2013, Acta Astron., 63, 21
Soszyński I., et al., 2016, Acta Astron., 66, 405
Specht D., Kerins E., Awiphan S., Robin A. C., 2020, MNRAS, 498, 2196
Sumi T., et al., 2011, Nature, 473, 349
Sumi T., et al., 2013, ApJ, 778, 150
Udalski A., 2003, Acta Astron., 53, 291
Udalski A., Pietrzynski G., Szymanski M., Kubiak M., Zebrun K., Soszynski I., Szewczyk O., Wyrzykowski L., 2003, Acta Astron., 53, 133
Udalski A., Szymański M. K., Szymański G., 2015a, Acta Astron., 65, 1
Udalski A., et al., 2015b, ApJ, 799, 237
Veras D., Raymond S. N., 2012, MNRAS, 421, L117
Vorobiev D., Ninkov Z., Caldwell D., Mochnacki S., 2018, preprint, (arXiv:1806.07430)
Wang D., Hogg D. W., Foreman-Mackey D., Schölkopf B., 2016, PASP, 128, 094503
Whitworth A. P., Zinnecker H., 2004, A&A, 427, 299
Witteborn F. C., Van Cleve J., Borucki W., Argabright V., Hascall P., 2011, in Proc. SPIE. p. 815117, doi:10.1117/12.892850
Yee J. C., et al., 2015, ApJ, 810, 155
Zang W., et al., 2018, PASP, 130, 104401
Zhu W., et al., 2017, PASP, 129, 104501

**Table 1.** Fitted $t_0$ and $t_{eff}$ of lens systems already detected by OGLE with $t_E < 10$ days.

| OGLE-2016-BLG- | OGLE location RA Dec (J2000) | | OGLE $t_0$ (JD)[1] | K2 | |
|---|---|---|---|---|---|
| | | | | $t_0$ (JD)[1] | $t_{eff}$ (days) |
| 0559 | 269.2141 | −28.4624 | 7505.29 | 7507.57 | 0.878 |
| 0770 | 269.6383 | −28.6178 | 7510.67 | 7510.60 | 0.408 |
| 0795 | 271.0011 | −28.1551 | 7512.63 | 7512.64 | 0.354 |
| 0863 | 268.7971 | −28.3588 | 7520.28 | 7519.70 | 0.408 |
| 0878 | 271.2549 | −27.7628 | 7525.64 | 7525.45 | 0.303 |
| 0885 | 269.8185 | −28.4355 | 7526.22 | 7525.72 | 0.844 |
| 0914 | 269.9763 | −28.6855 | 7542.28 | 7541.99 | 0.766 |
| 1231 | 269.2024 | −28.6552 | 7569.79 | 7569.66 | 0.256 |
| 1245 | 268.1876 | −29.0952 | 7568.52 | 7568.14 | 0.415 |

Notes: (1) Dates are given as truncated heliocentric Julian Dates, i.e., HJD − 2 450 000.

# APPENDIX

Table 1 lists the OGLE lenses recovered by the pipeline (recall that we intentionally only select short-timescale events, with longer events masked by smoothing functions). Four events with $t_E < 2$ days were undetected: OGLE-2016-BLG-0814 and -1043 because of their binary nature; 1041, which has a good-quality lightcurve but no event near the OGLE $t_0$; and -1097, which has a strongly asymmetric lightcurve and appears to be a mis-classified stellar flare. Note that -0559 is included in this list, as it would be modelled in the K2C9 data as a single-source, single-lens event, despite the fact that the OGLE photometry shows complex structure due to a third body.

Table 2 similarly lists the KMTNet lenses, including several long $t_E$ events (OGLE-2016-BLG-0768, -0971, -1145, and

**Table 2.** Fitted $t_0$ and $t_{eff}$ of lens systems already detected by KMTNet with $t_{eff} < 10$ days.

| KMT-2016-BLG- | KMTNet location RA Dec (J2000) | | KMTNet $t_0$ (JD)[1] | K2 $t_0$ (JD)[1] | $t_{eff}$ (days) |
|---|---|---|---|---|---|
| 0007 | 267.909 | −28.492 | 7510.85 | 7512.00 | 0.145 |
| 0025 | 268.810 | −28.615 | 7562.12 | 7561.71 | 1.098 |
| 0092 | 269.206 | −28.811 | 7544.56 | 7542.91 | 1.859 |
| 0095 | 269.122 | −28.978 | 7518.34 | 7517.03 | 1.504 |
| 0117 | 269.645 | −27.614 | 7506.50 | 7506.53 | 0.614 |
| 0128 | 270.932 | −27.859 | 7566.56 | 7566.61 | 0.309 |
| 0133 | 271.391 | −28.542 | 7562.78 | 7562.65 | 0.335 |
| 0138 | 271.039 | −28.176 | 7505.22 | 7505.68 | 0.611 |
| 0150 | 269.859 | −28.903 | 7559.86 | 7557.49 | 0.719 |
| 0162 | 269.336 | −28.308 | 7504.11 | 7504.12 | 0.216 |
| 0181 | 268.865 | −29.123 | 7503.44 | 7504.01 | 0.986 |
| 2554 | 268.895 | −28.477 | 7541.93 | 7539.59 | 0.519 |
| 2583 | 268.812 | −28.941 | 7507.22 | 7507.27 | 0.199 |

Notes: (1) Dates are given as truncated heliocentric Julian Dates, i.e., HJD − 2 450 000.

-1206) given short $t_{eff}$ by KMTNet, and the cataclysmic variable KMT-2016-BLG-0181. Several MOA events were also detected, but these are repeats of the tabulated OGLE and KMTNet events.

Figures 1 and 2 show *K2* photometry and our degenerate microlensing MCPM fits to known OGLE and KMTNet events that passed our selection pipeline, including (Figure 3) the cataclysmic variable KMT-2016-BLG-0181.

Several differences between $t_0$ from the ground-based data and *K2C9* $t_0$ are notable. Scarcity of ground-based data and correlated noise in *K2* can both cause fitting errors, which likely cause of the majority of differences. However, the differing *Kepler* and the Earth through the Einstein ring can also cause parallax effects with which the lens properties can be constrained. We visually examined all the known lightcurves and compared them to their ground-based discovery data. Significant differences in $t_0$ were found for OGLE-2016-BLG-1245, and KMT-2016-BLG-0090, -0128 (OGLE-2016-BLG-1206) and especially -2554, which may warrant further investigation.

Figure 4 also shows detailed lightcurves for our new candidates.

This paper has been typeset from a TeX/LaTeX file prepared by the author.

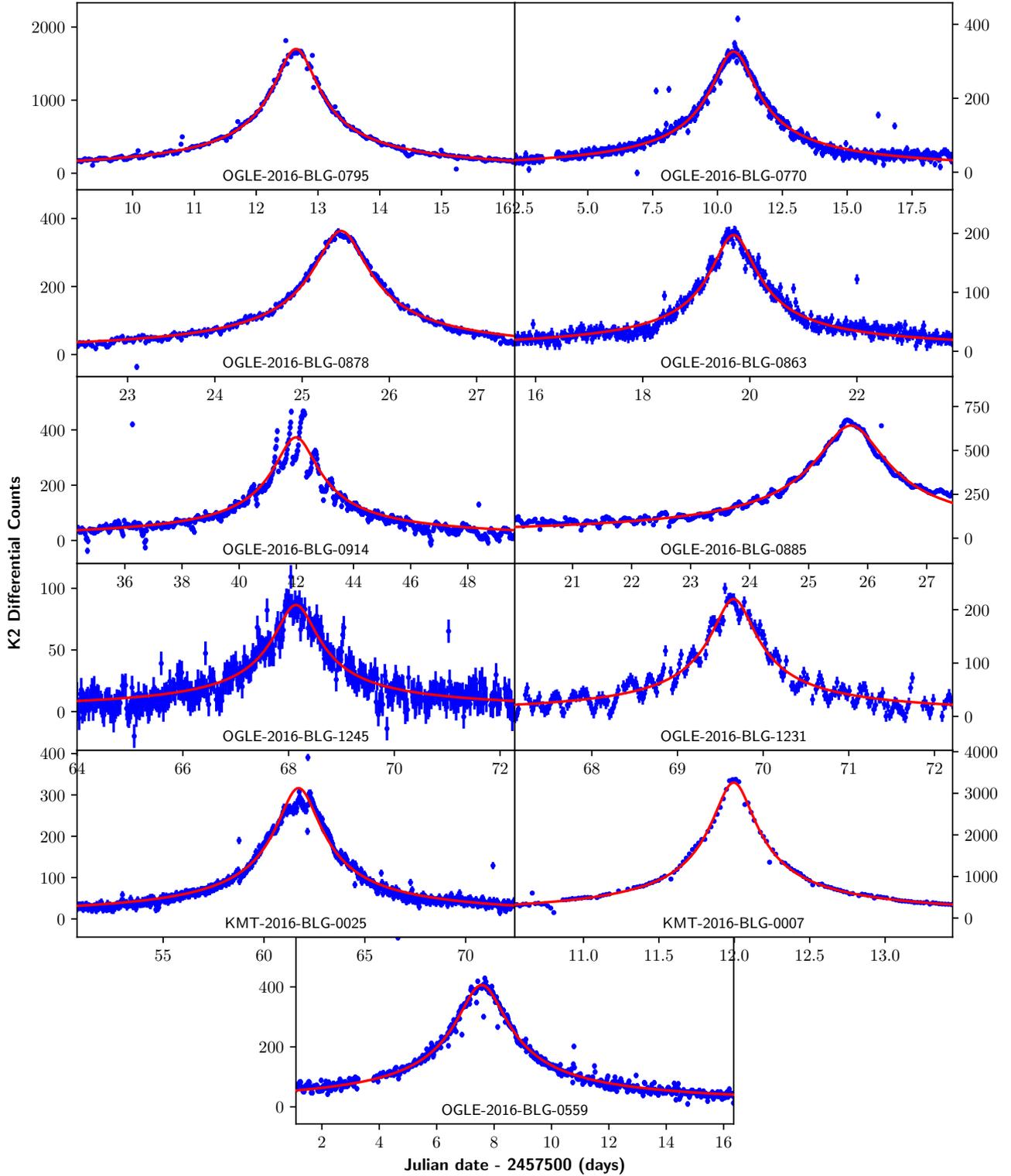

**Figure 1.** *K2*C9 lightcurves of known events that passed our selection pipeline cuts (blue points), and modelled with a degenerate microlening fit using MCPM (red lines).

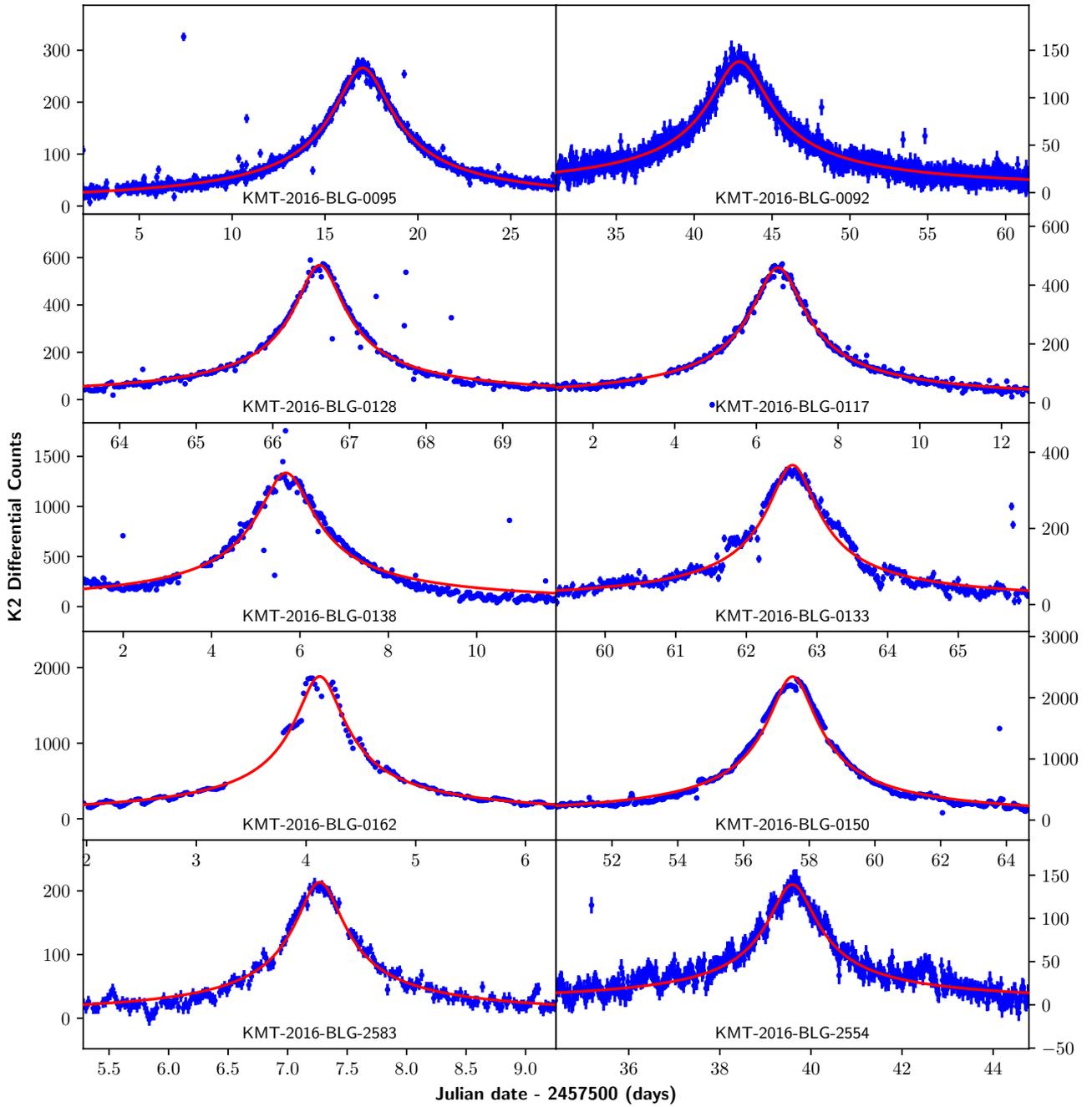

**Figure 2.** Figure 1 continued.

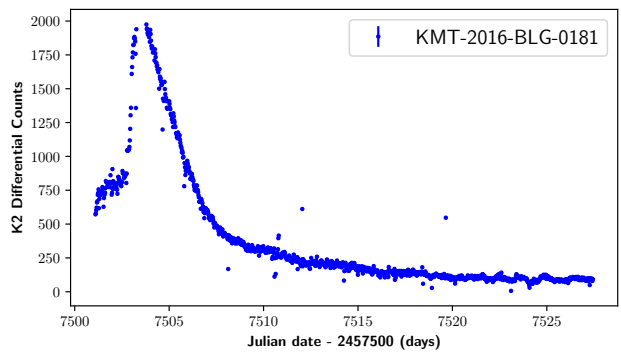

**Figure 3.** As Figure 1, but showing KMT-2016-BLG-0181, which is most likely a cataclysmic variable.

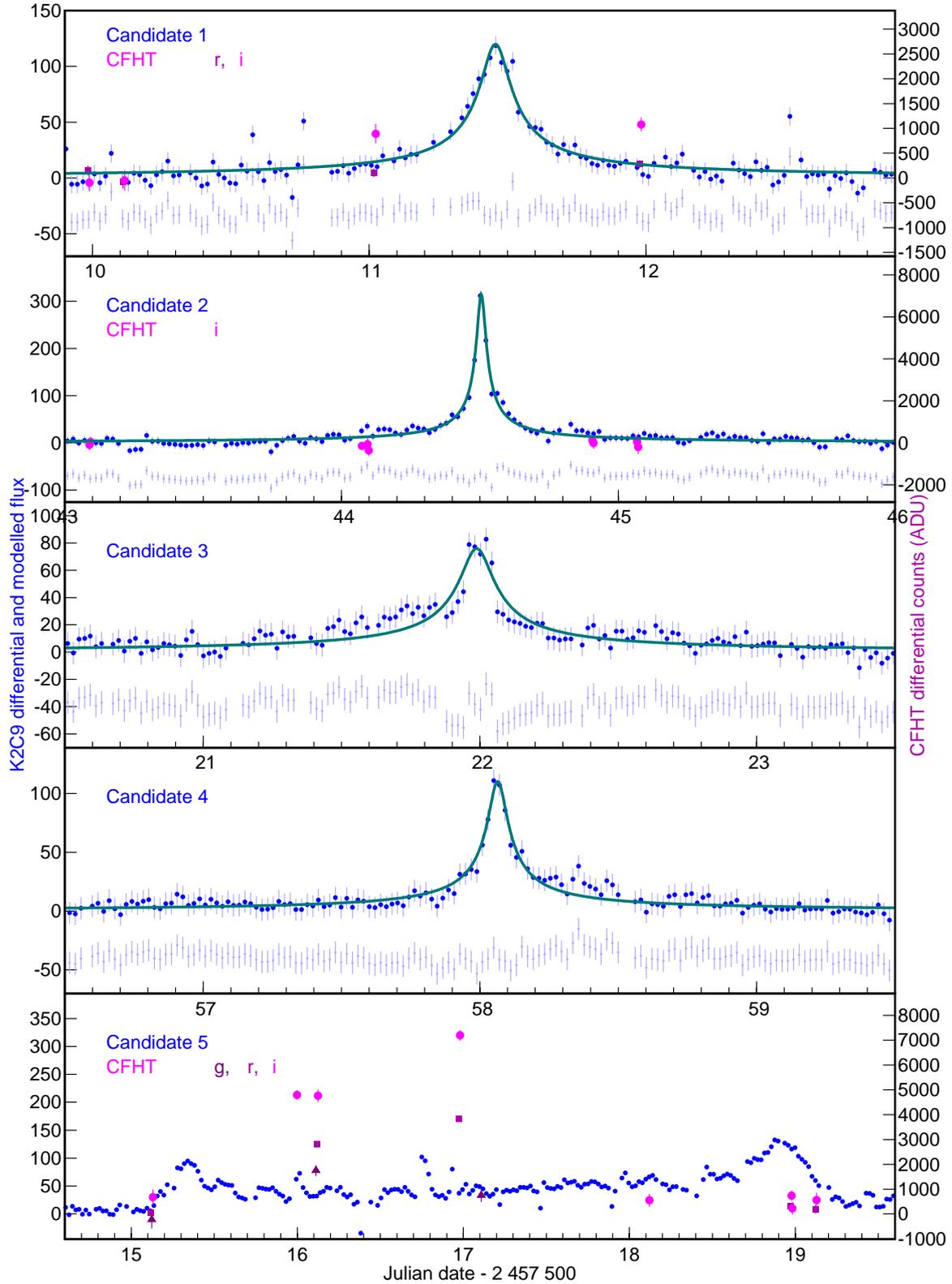

**Figure 4.** As Figure 11 in the main text, but detailing the region around the peak magnification.